\documentclass[aps,twocolumn,floatfix]{revtex4-1}
\usepackage{bm}
\usepackage{dcolumn}
\usepackage{graphicx}
\usepackage{amsmath}
\usepackage{amssymb} 
\usepackage{amsfonts}
\usepackage{float}
\usepackage{xcolor}
\usepackage[capitalise]{cleveref}

\begin{document}
\title{Diffusive density response of electrons in anisotropic multiband systems} 
\author{Jeonghyeon Suh$^{1}$}
\author{Sunghoon Kim$^{1}$}
\author{E. H. Hwang$^{2}$}
\email{euyheon@skku.edu}
\author{Hongki Min$^{1}$}
\email{hmin@snu.ac.kr}
\affiliation{$^{1}$ Department of Physics and Astronomy, Seoul National University, Seoul 08826, Korea}
\affiliation{$^{2}$ SKKU Advanced Institute of Nanotechnology and Department of Nano Engineering, Sungkyunkwan University, Suwon 16419, Korea}
\date{\today}

\begin{abstract}
We explicitly calculate the density-density response function with conserving vertex corrections for anisotropic multiband systems in the presence of impurities including long-range disorder. The direction-dependence of the vertex corrections is correctly considered to obtain the diffusion constant which is given by the combination of the componentwise transport relaxation times and velocities on the Fermi surface. We also investigate the diffusive density response of various anisotropic systems, propose some empirical rules for the corresponding diffusion constant, and demonstrate that it is crucial to consider the component-dependence of the transport relaxation times to correctly interpret the transport properties of anisotropic systems, especially various topological materials with a different power-law dispersion in each direction.
\end{abstract}

\maketitle

{\em Introduction.} ---
Recently, many anisotropic or multiband systems, such as black phosphorus with a tunable band gap \cite{Xia2019, Xia2014, Qiao2014, Tran2014, Kim2015, Baik2015, Li2016, Kim2017, Jang2019}, nodal line semimetals \cite{Fang2016, Fang2015, Huh2016, Han2017, Ahn2017-2, Rui2018, Chen2019, Shao2020}, and multi-Weyl semimetals \cite{Armitage2018, Fang2012, Ashby2014, Ahn2016, Ahn2017, Han2019, Nag2020, Ning2020, Fu2022}, have attracted much attention owing to their unique properties arising from their nodal structure with anisotropic nonlinear dispersion and associated chiral nature of the wave functions. It is essential to understand how the anisotropy and multiband nature are manifested in the physical properties of these systems.

The fundamental transport properties in the presence of impurities can be understood from the diffusive dynamics of current and density fluctuations in response to the external fields. The former corresponds to the current response characterized by the dc conductivity, whose form in anisotropic multiband systems has been obtained through the semiclassical Boltzmann transport theory \cite{Sorbello1974, Liu2016, Park2017, Kim2019} or many-body diagrammatic theory \cite{Kim2019}.
On the other hand, the latter corresponds to the density response characterized by the diffusion constant. In isotropic single-band systems, the density response takes the form
\begin{equation}
	\chi(\bm{q},\nu)\sim {1\over i\nu-{\cal D}q^2},
\end{equation}
which can be classically derived from the continuity equation $\frac{\partial \rho}{\partial t} + \nabla \cdot \bm{J} = 0$ and Fick's law $\bm{J} = - \mathcal{D} \nabla \rho$, where $\rho$, $\bm{J}$, and $\mathcal{D}$ are the number density, number current density, and diffusion constant, respectively. 
However, the diffusive density response of electrons in anisotropic multiband systems has not been exactly investigated in spite of its importance in understanding the corresponding diffusive transport. Thus, it is crucial to describe the density response correctly for anisotropic multiband systems in the presence of impurities.

In this paper, using the diagrammatic approach we develop a theory to correctly evaluate the vertex corrections to the density-density response function and corresponding diffusion constant in anisotropic multiband systems in the presence of disorders, including long-range disorder, within the low impurity density limit. We incorporate the direction-dependence of the vertex corrections originating from the chirality and long-range disorder of the systems, 
%additionally to the conventional derivations, 
and find that the diffusion constant is generally given by a nontrivial combination of the componentwise transport relaxation times $\tau^{(i)}$ and velocities $v^{(i)}$ ($i=x,y, \cdots$) on the Fermi surface, which satisfies the Einstein relation ensuring the consistency with the continuity equation.

We use our results to calculate the diffusion constants of anisotropic two-dimensional electron gas (2DEG), anisotropic graphene and few-layer black phosphorus (fBP) at the semi-Dirac transition point in the presence of long-range disorder for charged impurities. We demonstrate that the anisotropy of the diffusion constant (and also in the corresponding conductivity) strongly depends on the screening strength and deviates from the commonly expected anisotropy of the Fermi-velocity square, especially in highly anisotropic systems with a different power-law dispersion in each direction. Based on these observations, we propose some empirical rules for the anisotropy of the diffusion constant in anisotropic systems. We note that the anisotropy of the diffusion constant shows a significant difference from the one obtained neglecting the component-dependence of the transport relaxation time, %as in the isotropic systems.
indicating that the component-dependence of the transport relaxation time needs to be considered to correctly interpret the transport properties of anisotropic systems.

\begin{figure}[t]
\includegraphics[width=1.0\linewidth]{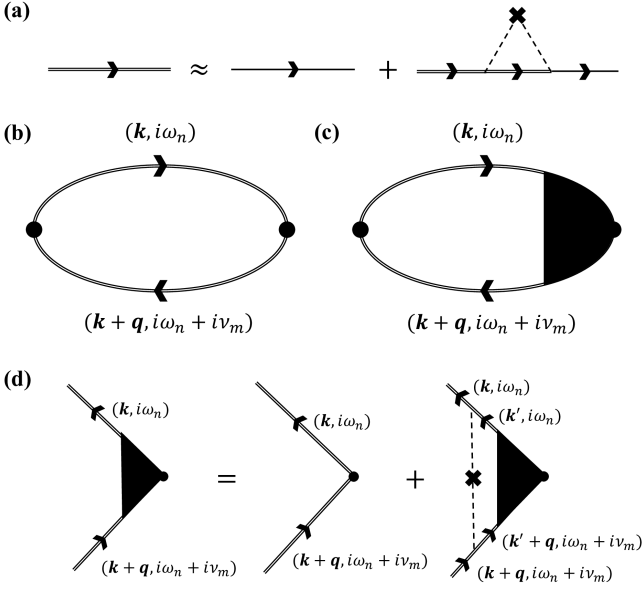}
\caption{
Feynman diagrams for (a) the disorder-averaged Green's function within the Born approximation, (b) the density-density response function without vertex corrections, (c) the density-density response function with vertex corrections and (d) the ladder approximation for the charge vertex.
} 
\label{fig:bubble_diagram}
\end{figure}

{\em Vertex corrections to the density-density response function.} ---
Within the ladder vertex corrections (Fig.~\ref{fig:bubble_diagram}), we establish the density-density response function of a disordered electron gas with the charge vertex $\Lambda_{0\alpha}$ for band $\alpha$ as follows:
\begin{align}
\label{eq:density_response_vertex_imaginary}
\chi(\bm{q},i\nu_m) & = \frac{\mathrm{g}}{\beta} \sum_{\alpha, i\omega_n} \int \frac{d^d k}{(2\pi)^d} \Lambda_{0\alpha} (\bm{k},i\omega_n;\bm{q},i\nu_m) \nonumber \\
&\times \mathcal{G}_\alpha (\bm{k},i\omega_n) \mathcal{G}_\alpha (\bm{k}+\bm{q},i\omega_n+i\nu_m),
\end{align}
where $\mathrm{g}$ is the spin/valley degeneracy factor, $\beta = 1/k_\textrm{B} T$, and $\omega_n$ and $\nu_m$ are fermionic and bosonic Matsubara frequencies, respectively. Here, $\mathcal{G}_\alpha (\bm{k},i\omega_n)$ is the disorder-averaged Green's function given by
\begin{equation}
\label{eq:disorder_green_function}
\mathcal{G}_\alpha (\bm{k}, i\omega_n) = \left[ i\omega_n - \xi_{\alpha,\bm{k}} - \Sigma_\alpha (\bm{k},i\omega_n) \right]^{-1},
\end{equation}
where $\xi_{\alpha, \bm{k}}$ is the energy measured from the Fermi energy at state $(\alpha, \bm{k})$ and $\Sigma_\alpha (\bm{k}, i\omega_n)$ is the electron self-energy due to impurity scattering. Here we assume a low temperature to ensure that the chemical potential can be approximated to the Fermi energy, and set $\hbar = 1$ for convenience.

Separating the charge vertex correction into two parts as $\Lambda_{0\alpha} = 1 + (\Lambda_{0\alpha} - 1)$, the density-density response function can be stated as $\chi = \chi_0 + \chi_1$. Here, $\chi_0$ is the density-density response function without vertex corrections, whose leading order term for impurities in the static long wavelength limit is given by $\chi_0 \approx -N(0)$ [see Sec.~\ref{sec:static_susceptibility} of the Supplemental Material (SM) \cite{SM}], where $N (\xi)$ is the density of states at energy $\xi$ measured from the Fermi energy. Then, the contribution of the vertex corrections is given by
\begin{align}
\label{eq:density_response_1_vertex_imaginary}
\chi_1(\bm{q},i\nu_m) & = \frac{\mathrm{g}}{\beta} \sum_{\alpha, i\omega_n} \int \frac{d^d k}{(2\pi)^d} \left[ \Lambda_{0\alpha} (\bm{k},i\omega_n;\bm{q},i\nu_m) - 1 \right] \nonumber \\
&\times \mathcal{G}_\alpha (\bm{k},i\omega_n) \mathcal{G}_\alpha (\bm{k}+\bm{q},i\omega_n+i\nu_m).
\end{align}

We begin with considering the Dyson equation for the charge vertex $\Lambda_{0\alpha}$ within the ladder approximation [Fig.~\ref{fig:bubble_diagram}(d)]
neglecting the quantum interference corrections:
\begin{align}
\label{eq:charge_vertex_dyson_equation_imaginary}
& \Lambda_{0\alpha} (\bm{k},i\omega_n ; \bm{q},i\nu_m) \nonumber \\ 
& = 1 + n_\textrm{imp} \sum_{\alpha^\prime} \int \frac{d^d k^\prime}{(2\pi)^d} \left| V_{\alpha, \bm{k} ;\alpha^\prime, \bm{k^\prime}} \right|^2 \Lambda_{0\alpha^\prime} (\bm{k^\prime},i\omega_n ; \bm{q},i\nu_m) \nonumber \\ 
& \quad\quad \times \mathcal{G}_{\alpha^\prime} (\bm{k^\prime},i\omega_n) \mathcal{G}_{\alpha^\prime} (\bm{k^\prime}+\bm{q},i\omega_n+i\nu_m),
\end{align}
where $n_\textrm{imp}$ is the impurity density and $V_{\alpha,\bm{k}; \alpha^\prime, \bm{k}^\prime}$ is the matrix element of the impurity potential between states $(\alpha,\bm{k})$ and $(\alpha^\prime, \bm{k}^\prime)$. In the long wavelength limit where $\bm{q} \rightarrow \bm{0}$ and in the low frequency-low impurity density limit where $\omega_n$ and $\Sigma_\alpha (\bm{k},i\omega_n)$ are negligible, Eq.~\eqref{eq:charge_vertex_dyson_equation_imaginary} transforms into
\begin{align}
	\label{eq:charge_vertex_dyson_equation_imaginary_alternative}
	&\Lambda_{0\alpha} (\bm{k},i\omega_n ; \bm{q},i\nu_m) - 1 \\
	&\approx \Theta_{n,m}\! \sum_{\alpha^\prime}\! \int\! \frac{d^d k^\prime}{(2\pi)^d} W_{\alpha,\bm{k} ; \alpha^\prime, \bm{k^\prime}} 
	\frac{\Lambda_{0\alpha^\prime} (\bm{k^\prime},i\omega_n ; \bm{q},i\nu_m)}{\nu_m + i \bm{q} \cdot \bm{v}_{\alpha^\prime, \bm{k^\prime}} + 1 / \tau_{\alpha^\prime, \bm{k^\prime}}^\textrm{qp}}, \nonumber
\end{align}
where $\Theta_{n,m}=1$ for $-\nu_m < \omega_n < 0$ and otherwise 0, $W_{\alpha,\bm{k} ; \alpha^\prime, \bm{k^\prime}} \equiv 2\pi n_\textrm{imp} |V_{\alpha,\bm{k} ; \alpha^\prime, \bm{k^\prime}}|^2 \delta(\xi_{\alpha,\bm{k}} - \xi_{\alpha^\prime,\bm{k^\prime}})$ is the transition rate from state $(\alpha,\bm{k})$ to $(\alpha^\prime,\bm{k^\prime})$, $\bm{v}_{\alpha, \bm{k}}$ is the velocity at $(\alpha, \bm{k})$, and $\tau_{\alpha, \bm{k}}^\textrm{qp}$ is the quasiparticle lifetime for $(\alpha,\bm{k})$ which is given up to the first-order Born approximation \cite{Flensberg2004} as 
\begin{equation}
	\label{eq:quasiparticle_lifetime}
	\frac{1}{\tau^\textrm{qp}_{\alpha,\bm{k}}} = \sum_{\alpha^\prime} \int \frac{d^d k^\prime}{(2\pi)^d} W_{\alpha,\bm{k};\alpha^\prime,\bm{k^\prime}}.
\end{equation}
For detailed derivations, see Sec.~\ref{sec:dyson_eqn_alter_derivation} of the SM \cite{SM}.

Similarly as in isotropic single-band systems \cite{Coleman2016}, the charge vertex with $\bm{q} = \bm{0}$ for $(\alpha,\bm{k})$ on the Fermi surface is given by (see Sec.~\ref{sec:ward_identity} of the SM \cite{SM})
\begin{equation}
\label{eq:charge_vertex_q=0}
\Lambda_{0\alpha} (\bm{k},i\omega_n;\bm{0},i\nu_m)  =1+ \frac{\Theta_{n,m}}{\nu_m \tau^\textrm{qp}_{\alpha,\bm{k}}}.
\end{equation}
Motivated from Eq.~\eqref{eq:charge_vertex_q=0}, we set an ansatz for the charge vertex as follows:
\begin{equation}
	\label{eq:charge_vertex_ansatz}
	\Lambda_{0\alpha} (\bm{k},i\omega_n;\bm{q},i\nu_m) =1+ \Theta_{n,m} \frac{1 - i \bm{q} \cdot \bm{l}_{\alpha,\bm{k}}}{\mathcal{V}_m (\bm{q},\nu_m) \tau_{\alpha, \bm{k}}^\textrm{qp}}
\end{equation}
for some $\bm{l}_{\alpha,\bm{k}}$ and $\mathcal{V}_m (\bm{q},\nu_m)$ satisfying $\mathcal{V}_m (\bm{0},\nu_m) = \nu_m$. The direction-dependence of the charge vertex from the coupling between $\bm{q}$ and $\bm{k}$, which has been conventionally neglected to obtain a solution of the Dyson equation in a closed form \cite{Brouwer2005, Coleman2016}, is considered up to linear order in $\bm{q}$ via $\bm{q} \cdot \bm{l}_{\alpha,\bm{k}}$ term. 

Inserting Eq.~\eqref{eq:charge_vertex_ansatz} to Eq.~\eqref{eq:charge_vertex_dyson_equation_imaginary_alternative} and expanding the right-hand side in powers of $\bm{q}$ and $\nu_m$, from the linear terms we obtain 
\begin{equation}
	\label{eq:l_ith_component}
	l^{(i)}_{\alpha,\bm{k}} = v_{\alpha,\bm{k}}^{(i)} \left( \tau_{\alpha,\bm{k}}^{(i)} - \tau_{\alpha,\bm{k}}^\textrm{qp} \right),
\end{equation}
where $v_{\alpha,\bm{k}}^{(i)}$ and $\tau_{\alpha,\bm{k}}^{(i)}$ are the $i$-th component of the velocity and transport relaxation time satisfying the following integral equation given by \cite{Sorbello1974,Liu2016,Park2017,Kim2019}
\begin{equation}
	\label{eq:transport_relaxation_time_integral_equation}
	1 = \sum_{\alpha^\prime} \int \frac{d^d k^\prime}{(2\pi)^d} W_{\alpha,\bm{k} ; \alpha^\prime, \bm{k^\prime}} \left(\tau_{\alpha,\bm{k}}^{(i)}-\frac{v^{(i)}_{\alpha^\prime,\bm{k^\prime}} }{v_{\alpha,\bm{k}}^{(i)}}\tau_{\alpha^\prime, \bm{k^\prime}}^{(i)}\right).
\end{equation}
As seen in Eq.~\eqref{eq:l_ith_component}, the $\bm{q} \cdot \bm{l}_{\alpha,\bm{k}}$ term added to the conventional derivations vanishes only if the quasiparticle lifetime and transport relaxation time coincide, which occurs for non-chiral systems with short-range disorder. Thus, we infer that the $\bm{q} \cdot \bm{l}_{\alpha,\bm{k}}$ term originates from the chirality and long-range disorder of the systems. On the other hand, from the quadratic terms averaged over the Fermi surface we obtain
\begin{equation}
	\label{eq:large_nu_m}
	\mathcal{V}_m (\bm{q},\nu_m) = \nu_m + \sum_{i,j} q_i q_j \mathcal{D}_{ij}  + O^3(\bm{q},\nu_m).
\end{equation}
Here $O^n(\bm{q},\nu_m)$ represents the subleading terms of $n$-th order or higher in $\bm{q}$ and $\nu_m$, and $\mathcal{D}_{ij}$ is the diffusion constant defined by
\begin{equation}
\label{eq:diffusion_constant}
\mathcal{D}_{ij} = \frac{1}{\tilde{N}(0)} \sum_{\alpha} \int \frac{d^d k}{(2\pi)^d} \delta(\xi_{\alpha,\bm{k}}) v_{\alpha,\bm{k}}^{(i)} v_{\alpha,\bm{k}}^{(j)} \tau_{\alpha,\bm{k}}^{(j)},
\end{equation}
where $\tilde{N} (\xi) \equiv N(\xi) / \mathrm{g}$ is the density of states per degeneracy at energy $\xi$. See Sec.~\ref{sec:charge_vertex_detailed_derivation} of the SM \cite{SM} for the detailed derivations of Eqs.~\eqref{eq:l_ith_component} and \eqref{eq:large_nu_m}. Note that the diffusion constant in Eq.~\eqref{eq:diffusion_constant} is symmetric with respect to the indices $i$ and $j$. Using Eq.~\eqref{eq:transport_relaxation_time_integral_equation}, Eq.~\eqref{eq:diffusion_constant} can be rewritten as
\begin{align}
	\label{eq:diffusion_constant_alternative}
	\mathcal{D}_{ij} & = \frac{1}{\tilde{N}(0)} \sum_{\alpha} \int \frac{d^d k}{(2\pi)^d} \delta(\xi_{\alpha,\bm{k}}) v_{\alpha,\bm{k}}^{(i)} v_{\alpha,\bm{k}}^{(j)} \tau_{\alpha,\bm{k}}^{(i)} \tau_{\alpha,\bm{k}}^{(j)} \left( \tau_{\alpha,\bm{k}}^{\textrm{qp}} \right)^{-1} \nonumber \\ & - \frac{1}{\tilde{N}(0)} \sum_{\alpha, \alpha^\prime} \int \frac{d^d k}{(2\pi)^d} \int \frac{d^d k^\prime}{(2\pi)^d} W_{\alpha,\bm{k};\alpha^\prime,\bm{k^\prime}} \nonumber \\ & \times \delta(\xi_{\alpha,\bm{k}}) v_{\alpha,\bm{k}}^{(i)} v_{\alpha^\prime,\bm{k^\prime}}^{(j)} \tau_{\alpha,\bm{k}}^{(i)} \tau_{\alpha^\prime,\bm{k^\prime}}^{(j)},
\end{align}
which clearly reflects the symmetry with respect to the indices $i$ and $j$. 

Repeating the process in Sec.~\ref{sec:dyson_eqn_alter_derivation} of the SM \cite{SM}, Eq.~\eqref{eq:density_response_1_vertex_imaginary} can be rewritten as
\begin{align}
	\label{eq:density_response_1_intermediate}
	\chi_1 (\bm{q},i\nu_m) & = \frac{2\pi \mathrm{g}}{\beta} \sum_{\alpha, i\omega_n} \Theta_{n,m} \int \frac{d^d k}{(2\pi)^d} \delta(\xi_{\alpha, \bm{k}}) \nonumber \\ & \times \frac{ \tau_{\alpha,\bm{k}}^\textrm{qp} \left[ \Lambda_{0\alpha} (\bm{k},i\omega_n ; \bm{q}, i\nu_m) - 1 \right]}{1 + \nu_m \tau_{\alpha, \bm{k}}^\textrm{qp} + i \bm{q} \cdot \bm{v}_{\alpha, \bm{k}} \tau_{\alpha, \bm{k}}^\textrm{qp}}.
\end{align}
Inserting Eq.~\eqref{eq:charge_vertex_ansatz} into Eq.~\eqref{eq:density_response_1_intermediate} and expanding the right-hand side, we finally obtain
\begin{align}
	\label{eq:density_response_1_expansion}
	\chi_1 (\bm{q},i\nu_m) & = N(0) \frac{\nu_m \left[ 1 + O^1 (\bm{q},\nu_m) \right]}{\nu_m + \sum_{i,j} q_i q_j \mathcal{D}_{ij} + O^3(\bm{q},\nu_m)}.
\end{align}
Here we have used $\frac{2\pi}{\beta} \sum_{i\omega_n} \Theta_{n,m} = \nu_m$. Therefore, up to leading order in $\bm{q}$ and $\nu_m$, $\chi(\bm{q},i\nu_m) = \chi_0 (\bm{q},i\nu_m) + \chi_1(\bm{q},i\nu_m)$ is given by
\begin{align}
	\chi(\bm{q},i\nu_m) \approx - N(0) \frac{\sum_{i,j} q_i q_j \mathcal{D}_{ij}}{\nu_m + \sum_{i,j} q_i q_j \mathcal{D}_{ij}}.
\end{align}
Through the analytic continuation $i\nu_m \rightarrow \nu + i0^+$, the retarded density-density response function is given by
\begin{equation}
	\label{eq:density_response_retarded}
	\chi^{(\textrm{R})} (\bm{q},\nu) = N(0) \frac{ \sum_{i,j} q_i q_j \mathcal{D}_{ij}}{i\nu - \sum_{i,j} q_i q_j \mathcal{D}_{ij}}.
\end{equation}
For the alternative derivations performing the frequency summation first, see Sec.~\ref{sec:alternative_derivation} of the SM \cite{SM}.

\begin{figure}[t]
\includegraphics[width=1\linewidth]{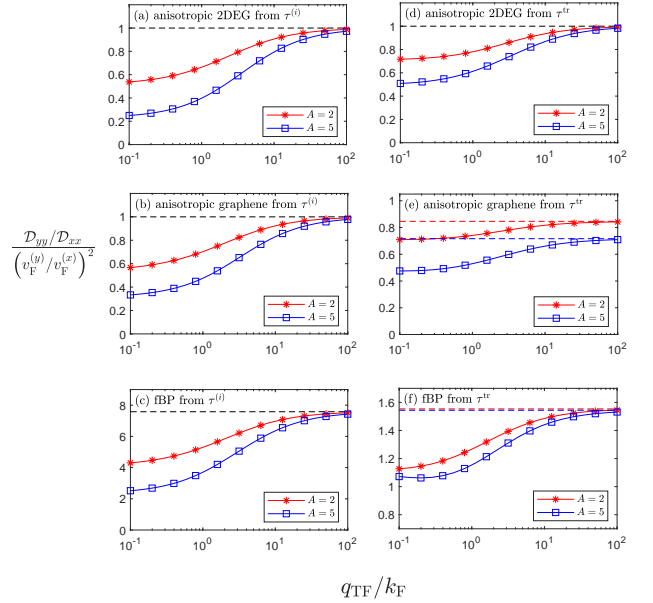}
\caption{
Anisotropy $\mathcal{D}_{yy} / \mathcal{D}_{xx}$ of the diffusion constant normalized by $(v_{\textrm{F}}^{(y)} / v_{\textrm{F}}^{(x)})^2$ as a function of the screening factor $Q$ for (a), (d) anisotropic 2DEG, (b), (e) anisotropic graphene and (c), (f) fBP at the semi-Dirac transition point, obtained from (a)-(c) Eq.~(\ref{eq:diffusion_constant}) considering the component-dependence of the transport relaxation time and from (d)-(f) Eq.~\eqref{eq:relaxation_time_approximation} neglecting the component-dependence of the transport relaxation time as in isotropic systems. Here $Q \equiv q_\textrm{TF} / k_\textrm{F}$ and $A \equiv k_{\textrm{F}}^{(x)} / k_{\textrm{F}}^{(y)}$. %is the anisotropy of the Fermi surface, where $k_{\textrm{F}}^{(i)}$ is the Fermi wavevector along the $i$-th direction. 
The values for the short-range disorder are represented by the dashed lines with the corresponding colors or by the black dashed lines if they coincide. %Note that for fBP at the semi-Dirac transition point $A = 2,5$ is estimated from the typical doping concentration $10^{12}$-$10^{13}$ $\mathrm{cm}^{-2}$ and realistic parameters calculated in Ref.~\cite{Jang2019}, whereas for the others $A$ is set equal to that of fBP for comparison. (see Sec.~\ref{sec:diffusion_constants_calculation} of the SM \cite{SM} for details.)
} 
\label{fig:diffusion_anisotropy}
\end{figure}

{\em Evaluation of the diffusion constants in anisotropic systems.} ---
We evaluate the diffusion constants in anisotropic 2DEG, anisotropic graphene and fBP at the semi-Dirac transition point for both short-range disorder and long-range disorder.
For the anisotropy factor $A=
k_{\textrm{F}}^{(x)} / k_{\textrm{F}}^{(y)}$ characterizing the anisotropy of the Fermi surface where $k_{\textrm{F}}^{(i)}$ is the Fermi wavevector along the $i$-th direction, we use $A = 2,5$ estimated from fBP at the semi-Dirac transition point with a typical doping concentration $n = 10^{12}$-$10^{13}$ $\mathrm{cm}^{-2}$, whereas for anisotropic 2DEG and anisotropic graphene, we use the same $A$ for comparison. See Sec.~\ref{sec:diffusion_constants_calculation} of the SM \cite{SM} for details. 

In anisotropic 2DEG and anisotropic graphene, $\mathcal{D}_{yy} / \mathcal{D}_{xx}$ is equal to the commonly expected $(v_{\textrm{F}}^{(y)} / v_{\textrm{F}}^{(x)})^2$ for short-range disorder, where $v_{\textrm{F}}^{(i)}$ is the Fermi velocity along the $i$-th direction, whereas for long-range disorder it deviates from $(v_{\textrm{F}}^{(y)} / v_{\textrm{F}}^{(x)})^2$ and depends on the screening factor $Q \equiv q_\textrm{TF} / k_\textrm{F}$ characterizing the screening strength, where $q_{\rm TF}$ is the Thomas-Fermi wavevector and $k_{\rm F}$ is the effective Fermi wavevector. In the strong screening limit, the result eventually approaches that of the short-range disorder [Figs.~\ref{fig:diffusion_anisotropy}(a) and \ref{fig:diffusion_anisotropy}(b)].
In fBP at the semi-Dirac transition point where the energy dispersion is quadratic/linear along the zigzag ($x$)/armchair ($y$) direction with different power-laws depending on the direction, $\mathcal{D}_{yy} / \mathcal{D}_{xx}$ becomes $7.6 (v_{\textrm{F}}^{(y)} / v_{\textrm{F}}^{(x)})^2$ differing from $(v_{\textrm{F}}^{(y)} / v_{\textrm{F}}^{(x)})^2$ even for short-range disorder. For long-range disorder, it increases with the screening strength, approaching the short-range result in the strong screening limit [Fig.~\ref{fig:diffusion_anisotropy}(c)]. Note that the dependence on the screening strength becomes larger as the anisotropy of the system increases for all cases [Figs.~\ref{fig:diffusion_anisotropy}(a), \ref{fig:diffusion_anisotropy}(b), and \ref{fig:diffusion_anisotropy}(c)]. %(Detailed dependence on the anisotropy is illustrated in Sec.~\ref{sec:diffusion_constants_calculation} of the SM \cite{SM})
For the detailed derivations and numerical calculations, see Sec.~\ref{sec:diffusion_constants_calculation} of the SM \cite{SM}.

When the system has the same power-law dispersion in each direction as in anisotropic 2DEG and anisotropic graphene, for short-range disorder $\tau_{\bm{k}}^{(i)}$ becomes the same for each component $i$ and independent of the direction of $\bm{k}$ that $\mathcal{D}_{yy}/\mathcal{D}_{xx}=(v_{\textrm{F}}^{(y)} / v_{\textrm{F}}^{(x)})^2$. For long-range disorder, $\tau_{\bm{k}}^{(i)}$ has a dependence not only on the direction of $\bm{k}$ but also on $i$ that the deviation of $\mathcal{D}_{yy}/\mathcal{D}_{xx}$ from $(v_{\textrm{F}}^{(y)} / v_{\textrm{F}}^{(x)})^2$ increases as the screening becomes weaker. When the system has a different power-law dispersion in each direction as in fBP at the semi-Dirac transition point, $\tau_{\bm{k}}^{(i)}$ has a dependence not only on the direction of $\bm{k}$ but also on the component $i$ even for short-range disorder, yielding a significant deviation of $\mathcal{D}_{yy}/\mathcal{D}_{xx}$ from $(v_{\textrm{F}}^{(y)} / v_{\textrm{F}}^{(x)})^2$. In both cases, the deviation in anisotropy arising from $\tau_{\bm{k}}^{(i)}$ shows a stronger dependence on the screening compared to that obtained from
\begin{equation}
	\label{eq:relaxation_time_approximation}
	{1\over \tau_{\bm k}^\textrm{tr}}=\int {d^d k' \over (2\pi)^d} W_{\bm{k};\bm{k}'} (1-\hat{\bm{k}}\cdot\hat{\bm{k}}')
\end{equation}
neglecting the dependence on the component $i$ as in isotropic systems [Figs.~\ref{fig:diffusion_anisotropy}(d), \ref{fig:diffusion_anisotropy}(e) and \ref{fig:diffusion_anisotropy}(f)].
From these observations, we find that the componentwise transport relaxation time should be considered to correctly interpret the transport properties of anisotropic systems, especially when dealing with highly anisotropic systems with a different power-law dispersion in each direction, even in the strong screening limit. %rather than approximate it to be isotropic as in the system analogous to anisotropic electron gas.

Furthermore, from the Einstein relation in anisotropic multiband systems \cite{Sorbello1974,Liu2016,Park2017,Kim2019}, the dc conductivity is given by
\begin{equation}
\label{eq:Einstein_relation}
	\sigma_{ij} (\bm{q} \rightarrow \bm{0}) = e^2 N(0) \mathcal{D}_{ij},
\end{equation}
thus we have $\sigma_{yy}/\sigma_{xx}=\mathcal{D}_{yy}/\mathcal{D}_{xx}$. Consequently, the anisotropy of the conductivity also shows a significant difference from that estimated neglecting the component-dependence of transport relaxation time for long-range disorder, and even for short-range disorder when the system has a different power-law dispersion in each direction.

{\em Discussion.} ---
In $d$-dimensional isotropic single-band systems, the diffusion constant in Eq.~\eqref{eq:diffusion_constant} reduces to
\begin{equation}
\label{eq:diffusion_constant_isotropic}
\mathcal{D} = \frac{v_{\rm{F}}^2 \tau^\textrm{tr}}{d},
\end{equation}
which has the same form appearing in the Einstein relation. However, the conventional many-body diagrammatic approach considering the vertex corrections to the density-density response function gives the diffusion constant to be \cite{Brouwer2005,Coleman2016}
\begin{equation}
\label{eq:diffusion_constant_isotropic_known}
\mathcal{D} = \frac{v_{\rm{F}}^2 \tau^\textrm{qp}}{d},
\end{equation}
where $\tau^\textrm{qp}$ is the quasiparticle lifetime. The difference between the conventional approach and our results originates from the additional $\bm{q} \cdot \bm{l}_{\bm{k}}$ term in Eq.~\eqref{eq:charge_vertex_ansatz}, which is the only direction-dependence on $\bm{q}$ for isotropic systems. As mentioned, the conventional approach in isotropic single-band systems neglects the direction-dependence of the charge vertex to obtain a solution of the Dyson equation in a closed form. However, the Dyson equation in Eq.~\eqref{eq:charge_vertex_dyson_equation_imaginary} actually depends on the direction of $\bm{q}$ through the $\bm{k}$-dependence in $W_{\alpha,\bm{k};\alpha^\prime,\bm{k^\prime}}$ when the system has chirality or long-range disorder.
%in the presence of chirality or long-range disorder. 
We correctly considered this direction-dependence in the Dyson equation and obtain the corresponding solutions up to linear order in $\bm{q}$, and to quadratic order in $\bm{q}$ averaged over the Fermi surface, respectively.

Furthermore, the diffusion constant given by Eq.~\eqref{eq:diffusion_constant} correctly describes the diffusive dynamics. From the continuity equation $\frac{\partial \rho}{\partial t} + \nabla \cdot \bm{J} = 0$, the density-density response function and conductivity are related as $i\nu e^2 \chi(\bm{q},\nu) + \sum_{i,j} \sigma_{ij} q_i q_j = 0$. Thus, using $\chi(\bm{q} \rightarrow \bm{0},\nu) \approx N(0) \sum_{i,j} q_i q_j \mathcal{D}_{ij} / i\nu$ from Eq.~\eqref{eq:density_response_retarded}, we can reproduce the Einstein relation in Eq.~\eqref{eq:Einstein_relation}.

In disordered systems, the density-density response function has the diffusion pole presenting a pronounced peak at low frequencies in the density fluctuation spectrum, which affects the quasiparticle properties of a disordered electron liquid \cite{Giuliani2005}. In anisotropic multiband systems, the density-density response function is given by Eq.~\eqref{eq:density_response_retarded} characterized by the diffusion pole structure, thus the diffusion pole occurs at $\nu = -i \sum_{i,j} q_i q_j \mathcal{D}_{ij}$. Since the diffusion constant given by Eq.~\eqref{eq:diffusion_constant} is anisotropic in general, the diffusion pole occurring due to disorder may affect the quasiparticle properties anisotropically. Thus, studying the anisotropy of the diffusion constant correctly considering the component-dependence of the transport relaxation time is important to understand the effect of disorder in anisotropic multiband systems.

In summary, using a many-body diagrammatic approach, we develop a theory for the vertex corrections to the density-density response function and find the corresponding diffusion constant in anisotropic multiband systems.
We fully incorporate the direction-dependence of the charge vertex, especially the one from the chirality and long-range disorder of the systems, and find that the diffusion constant obtained in this many-body diagrammatic approach is associated with the componentwise transport relaxation time rather than the quasiparticle lifetime. This nontrivial result correctly describes the diffusive dynamics of anisotropic multiband systems, consistent with the continuity equation.
Furthermore, we calculate the diffusion constants of various anisotropic systems in the presence of the long-range disorder for charged impurities and find that the inclusion of the component-dependent transport relaxation time is crucial to correctly describe the transport properties of anisotropic systems.

\acknowledgments
This work was supported by the National Research Foundation of Korea (NRF) grant funded by the Korea government (MSIT) (Grant No. 2018R1A2B6007837) and Creative-Pioneering Researchers Program through Seoul National University (SNU).  E. H. acknowledges support from Korea NRF (Grant No. 2021R1A2C1012176).

\clearpage
\widetext
\setcounter{section}{0}
\setcounter{equation}{0}
\setcounter{figure}{0}
\setcounter{table}{0}
\renewcommand{\theequation}{S\arabic{equation}}
\renewcommand\thefigure{S\arabic{figure}} 
\setcounter{page}{1}
\begin{center}
\textbf{\large Supplemental Material: \\
Diffusive density response of electrons in anisotropic multiband systems}
\end{center}
\begin{center}
\author{Jeonghyeon Seo$^{1}$}
\author{Sunghoon Kim$^{1}$}
\author{E. H. Hwang$^{2}$}
\author{Hongki Min$^{1}$}
\email{hmin@snu.ac.kr}
\affiliation{$^{1}$ Department of Physics and Astronomy, Seoul National University, Seoul 08826, Korea}
\affiliation{$^{2}$ SKKU Advanced Institute of Nanotechnology and Department of Physics, Sungkyunkwan University, Suwon 16419, Korea}
\end{center}

\section{Static susceptibility in the long wavelength limit}
\label{sec:static_susceptibility}
\begin{figure}[htb]
	\includegraphics[width=.5\linewidth]{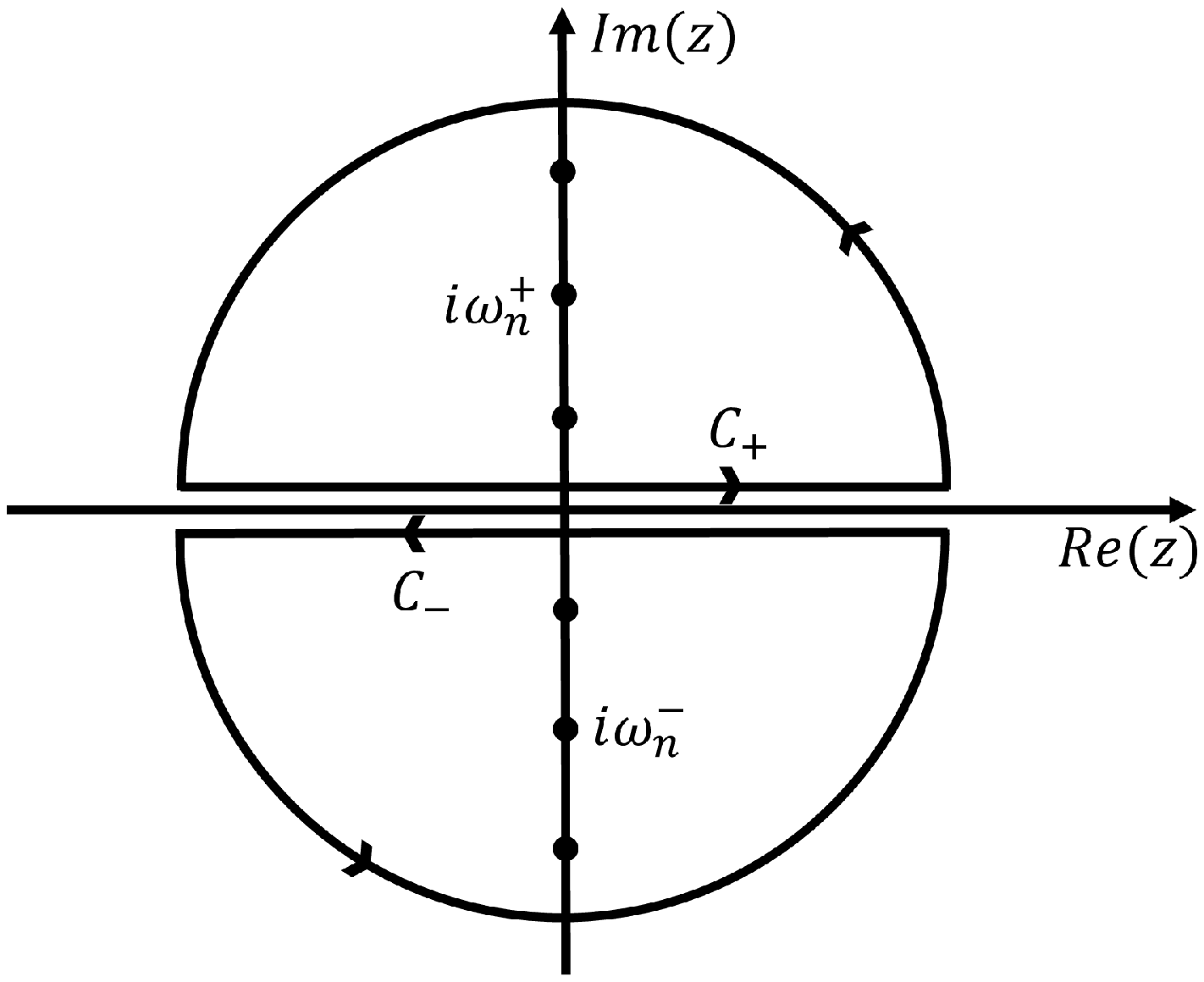}
	\caption{Contour $C_{\pm}$ used in Eq.~\eqref{eq:static_susceptibility_intermediate}. Note that dots on the upper and lower half plane represent $i\omega_n^{+}$ and $i\omega_n^{-}$, respectively.} 
	\label{fig:contour_C_pm}
\end{figure}

In the long wavelength limit, the static susceptibility is given by
\begin{equation}
	\label{eq:static_susceptibility_green_function}
	\chi_0 (\bm{0},0) = \frac{\rm{g}}{\beta} \sum_{\alpha,i\omega_n} \int \frac{d^d k}{(2\pi)^d} \left[ \mathcal{G}_\alpha (\bm{k}, i\omega_n) \right]^2 ,
\end{equation}
where the disorder-averaged Green's function $\mathcal{G}_\alpha (\bm{k}, i\omega_n)$ for band $\alpha$ is given by
\begin{equation}
	\mathcal{G}_\alpha (\bm{k}, i\omega_n) = \left[ i\omega_n - \xi_{\alpha,\bm{k}} - \Sigma_\alpha (\bm{k}, i\omega_n) \right]^{-1}.
\end{equation}
Separating $\sum_{i\omega_n} = \sum_{i\omega_n^+} + \sum_{i\omega_n^-}$, where $i\omega_n^+$ and $i\omega_n^-$ represent the Matsubara frequencies in the upper and lower half of the complex plane, respectively, the residue theorem transforms Eq.~\eqref{eq:static_susceptibility_green_function} into
\begin{align}
	\label{eq:static_susceptibility_intermediate}
	\chi_0(\bm{0},0) & = - \mathrm{g} \sum_\alpha \int \frac{d^d k}{(2\pi)^d} \left( \oint_{C_{+}} + \oint_{C_{-}} \right) \frac{dz}{2\pi i} f^{(0)} (z) \left[ \mathcal{G}_\alpha (\bm{k}, z) \right]^2 \nonumber \\ 
	& = - \mathrm{g} \sum_\alpha \int \frac{d^d k}{(2\pi)^d} \!\! \int \frac{d\xi}{2\pi i} f^{(0)} (\xi)
	\left\{ \left[ \mathcal{G}_\alpha^\textrm{R} (\bm{k}, \xi) \right]^2 - \left[ \mathcal{G}_\alpha^\textrm{A} (\bm{k}, \xi) \right]^2 \right\} \\ 
	& = \mathrm{g} \sum_\alpha \int \frac{d^d k}{(2\pi)^d} \int \frac{d\xi}{2\pi i} S^{(0)} (\xi) \left[ \mathcal{G}_\alpha^\textrm{R} (\bm{k}, \xi) - \mathcal{G}_\alpha^\textrm{A} (\bm{k}, \xi)\right], \nonumber
\end{align}
where $f^{(0)} (\xi) \equiv (e^{\beta\xi} + 1)^{-1}$ is the Fermi-Dirac distribution function, $S^{(0)} (\xi) \equiv - \frac{df^{(0)} (\xi)}{d\xi}$, the contour $C_{\pm}$ is represented in Fig.~\ref{fig:contour_C_pm}, and the superscripts $\textrm{A}$ and $\textrm{R}$ represent the advanced and retarded functions specified by $i\omega_n \rightarrow \xi \mp i0^+$ ensuring that $\mathcal{G}_\alpha^\textrm{A} (\bm{k}, \xi) \equiv \mathcal{G}_\alpha (\bm{k}, \xi - i0^+ )$ and $\mathcal{G}_\alpha^\textrm{R} (\bm{k}, \xi) \equiv \mathcal{G}_\alpha (\bm{k}, \xi + i0^+ )$, respectively. Note that integration by parts is used when analyzing the last equality, assuming that the self-energy varies negligibly slower than $\xi$.
Because the real part of the self-energy can be integrated into the definition of the chemical potential \cite{S_Flensberg2004, S_Brouwer2005}, Eq.~\eqref{eq:static_susceptibility_intermediate} is rewritten as
\begin{equation}
	\label{eq:static_susceptibility_final}
	\chi_0(\bm{0},0) = - \mathrm{g} \sum_\alpha \int \frac{d^d k}{(2\pi)^d} \int \frac{d\xi}{\pi} S^{(0)} (\xi) \frac{\Delta_{\alpha,\bm{k}} (\xi)}{(\xi - \xi_{\alpha,\bm{k}})^2 + \Delta_{\alpha,\bm{k}}^2 (\xi)},
\end{equation}
where $\Delta_{\alpha, \bm{k}} (\xi) \equiv \operatorname{Im} \Sigma_\alpha^\textrm{A} (\bm{k},\xi) > 0$. Here, $\Sigma_\alpha^\textrm{R} (\bm{k},\xi) = \Sigma_\alpha^{\textrm{A}\star} (\bm{k},\xi)$ is used, where $\star$ represents the complex conjugation. Assuming a low impurity density, $\Delta_{\alpha,\bm{k}} (\xi)$ is much smaller than the typical energy scale, resulting in
\begin{equation}
	\label{eq:lorentzian_delta_function}
	\frac{\Delta_{\alpha,\bm{k}} (\xi)}{(\xi - \xi_{\alpha,\bm{k}})^2 + \Delta_{\alpha,\bm{k}}^2 (\xi)} \approx \pi \delta(\xi - \xi_{\alpha,\bm{k}}).
\end{equation}
Thus, inserting Eq.~\eqref{eq:lorentzian_delta_function} into Eq.~\eqref{eq:static_susceptibility_final} considering the low temperature approximation $S^{(0)} (\xi) \approx \delta(\xi)$, we obtain
\begin{equation}
	\chi_0(\bm{q},i\nu_m) \approx \chi_0(\bm{0},0) \approx - N (0),
\end{equation}
where $N(\xi)$ is the density of states per unit volume at energy $\xi$.

\section{Detailed derivations of the alternative form of Dyson equation}
\label{sec:dyson_eqn_alter_derivation}

Since $\mathcal{G}_\alpha (\bm{k},i\omega_n) \mathcal{G}_\alpha (\bm{k}+\bm{q},i\omega_n+i\nu_m)$ can be rewritten as
\begin{equation}
	\label{eq:green_function_product_imaginary_alternative_SM}
	\mathcal{G}_\alpha (\bm{k},i\omega_n) \mathcal{G}_\alpha (\bm{k} + \bm{q},i\omega_n + i\nu_m) = \frac{\mathcal{G}_\alpha (\bm{k},i\omega_n) - \mathcal{G}_\alpha (\bm{k} + \bm{q},i\omega_n + i\nu_m)}{\mathcal{G}_\alpha^{-1} (\bm{k} + \bm{q},i\omega_n + i\nu_m) - \mathcal{G}_\alpha^{-1} (\bm{k},i\omega_n)},
\end{equation}
Eq.~\eqref{eq:charge_vertex_dyson_equation_imaginary} transforms as follows in $\bm{q} \rightarrow \bm{0}$ limit:
\begin{align}
	\label{eq:charge_vertex_dyson_equation_imaginary_intermediate_SM}
	\Lambda_{0\alpha} (\bm{k},i\omega_n ; \bm{q},i\nu_m) - 1 & = n_\textrm{imp} \int d\xi \left[ \mathcal{G} (\xi, i\omega_n) - \mathcal{G} (\xi, i\omega_n + i\nu_m) \right] \sum_{\alpha^\prime} \int \frac{d^d k^\prime}{(2\pi)^d} \left| V_{\alpha, \bm{k} ;\alpha^\prime, \bm{k^\prime}} \right|^2 \delta(\xi - \xi_{\alpha^\prime, \bm{k^\prime}}) \nonumber \\ & \times \frac{\Lambda_{0\alpha^\prime} (\bm{k^\prime},i\omega_n ; \bm{q},i\nu_m)}{\mathcal{G}_{\alpha^\prime}^{-1} (\bm{k^\prime} + \bm{q},i\omega_n + i\nu_m) - \mathcal{G}_{\alpha^\prime}^{-1} (\bm{k^\prime},i\omega_n)},
\end{align}
where $\mathcal{G} (\xi, i\omega_n) \equiv \mathcal{G}_\alpha (\bm{k},i\omega_n)$ for $\xi = \xi_{\alpha,\bm{k}}$. Because the Green's function defined by Eq.~\eqref{eq:disorder_green_function} has a large peak near the Fermi surface in the low frequency-low impurity density limit where $\omega_n$ and $\Sigma_\alpha (\bm{k},i\omega_n)$ are negligible, we can set $\xi\approx 0$ in the delta function. Now, we restrict the state $(\alpha,\bm{k})$ on the Fermi surface ensuring that $\xi_{\alpha,\bm{k}} = 0$. Using $\int d\xi \mathcal{G} (\xi, i\omega_n) \approx -i\pi {\rm sgn}(\omega_n)$ in the low impurity density limit, Eq.~\eqref{eq:charge_vertex_dyson_equation_imaginary_intermediate_SM} transforms into Eq.~\eqref{eq:charge_vertex_dyson_equation_imaginary_alternative}. Here for $(\alpha,\bm{k})$ on the Fermi surface we have used 
\begin{equation}
	\label{eq:green_function_product_imaginary_denominator}
	\mathcal{G}_\alpha^{-1} (\bm{k} + \bm{q}, i\omega_n + i\nu_m) - \mathcal{G}_\alpha^{-1} (\bm{k}, i\omega_n)	= i\nu_m - \bm{q} \cdot \bm{v}_{\alpha, \bm{k}} + \frac{i}{\tau^\textrm{qp}_{\alpha,\bm{k}}}
\end{equation}
in the range where $\Theta_{n,m} = 1$, which can be derived from the 
the Ward identity \cite{S_Schrieffer1964} in the low frequency-long wavelength limit (see Sec.~\ref{sec:ward_identity} for details).

\section{Ward identity and evaluation of the charge/current vertex at $\bm{q}=0$}
\label{sec:ward_identity}
In the low frequency-long wavelength limit, the Ward identity is given by \cite{S_Schrieffer1964}
\begin{equation}
	\label{eq:ward_identity}
	\mathcal{G}_{\alpha}^{-1} (\bm{k} + \bm{q},i\omega_n + i\nu_m) - \mathcal{G}_{\alpha}^{-1} (\bm{k},i\omega_n) = i\nu_m \Lambda_{0\alpha} (\bm{k},i\omega_n ; \bm{q}, i\nu_m) - \sum_i q_i v_{\alpha,\bm{k}}^{(i)} \Lambda_\alpha^{(i)} (\bm{k},i\omega_n ; \bm{q}, i\nu_m),
\end{equation}
where $\Lambda_\alpha^{(i)}$ is the current vertex defined by
\begin{align}
	\label{eq:current_vertex_dyson_equation_imaginary}
	& v_{\alpha,\bm{k}}^{(i)} \Lambda_{\alpha}^{(i)} (\bm{k},i\omega_n ; \bm{q},i\nu_m) - v_{\alpha,\bm{k}}^{(i)} \\ & = n_\textrm{imp} \sum_{\alpha^\prime} \int \frac{d^d k^\prime}{(2\pi)^d} \left| V_{\alpha, \bm{k} ;\alpha^\prime, \bm{k^\prime}} \right|^2 v_{\alpha^\prime,\bm{k^\prime}}^{(i)} \Lambda_{\alpha^\prime}^{(i)} (\bm{k^\prime},i\omega_n ; \bm{q},i\nu_m) \mathcal{G}_{\alpha^\prime} (\bm{k^\prime},i\omega_n) \mathcal{G}_{\alpha^\prime} (\bm{k^\prime}+\bm{q},i\omega_n+i\nu_m). \nonumber
\end{align}
Using Eq.~\eqref{eq:ward_identity} instead of \eqref{eq:green_function_product_imaginary_denominator}, Eq.~\eqref{eq:charge_vertex_dyson_equation_imaginary_alternative} is rewritten as
\begin{align}
	\label{eq:charge_vertex_dyson_equation_imaginary_alternative_pre}
	& \Lambda_{0\alpha} (\bm{k},i\omega_n ; \bm{q},i\nu_m) - 1 \\ & = \Theta_{n,m} \sum_{\alpha^\prime} \int \frac{d^d k^\prime}{(2\pi)^d} W_{\alpha,\bm{k} ; \alpha^\prime, \bm{k^\prime}} \frac{\Lambda_{0\alpha^\prime} (\bm{k^\prime},i\omega_n ; \bm{q},i\nu_m)}{\nu_m \Lambda_{0\alpha^\prime} (\bm{k^\prime},i\omega_n ; \bm{q}, i\nu_m) + i\sum_i q_i v_{\alpha^\prime,\bm{k^\prime}}^{(i)} \Lambda_{\alpha^\prime}^{(i)} (\bm{k^\prime},i\omega_n ; \bm{q}, i\nu_m)}. \nonumber
\end{align}
Similarly, repeating the process used in Eqs.~\eqref{eq:charge_vertex_dyson_equation_imaginary} to \eqref{eq:charge_vertex_dyson_equation_imaginary_alternative} and using Eq.~\eqref{eq:ward_identity} instead of \eqref{eq:green_function_product_imaginary_denominator}, Eq.~\eqref{eq:current_vertex_dyson_equation_imaginary} transforms into
\begin{align}
	\label{eq:current_vertex_dyson_equation_imaginary_alternative_pre}
	& v_{\alpha,\bm{k}}^{(i)} \Lambda_{\alpha}^{(i)} (\bm{k},i\omega_n ; \bm{q},i\nu_m) - v_{\alpha,\bm{k}}^{(i)} \\ & = \Theta_{n,m} \sum_{\alpha^\prime} \int \frac{d^d k^\prime}{(2\pi)^d} W_{\alpha,\bm{k} ; \alpha^\prime, \bm{k^\prime}} \frac{v_{\alpha^\prime,\bm{k^\prime}}^{(i)} \Lambda_{\alpha^\prime}^{(i)} (\bm{k^\prime},i\omega_n ; \bm{q},i\nu_m)}{\nu_m \Lambda_{0\alpha^\prime} (\bm{k^\prime},i\omega_n ; \bm{q}, i\nu_m) + i\sum_i q_i v_{\alpha^\prime,\bm{k^\prime}}^{(i)} \Lambda_{\alpha^\prime}^{(i)} (\bm{k^\prime},i\omega_n ; \bm{q}, i\nu_m)}. \nonumber
\end{align}
Thus, combining Eqs.~\eqref{eq:charge_vertex_dyson_equation_imaginary_alternative_pre} and \eqref{eq:current_vertex_dyson_equation_imaginary_alternative_pre} with the aid of Eq.~\eqref{eq:quasiparticle_lifetime}, we obtain
\begin{equation}
	\label{eq:charge_current_dyson_equation_sum}
	\nu_m \left[ \Lambda_{0\alpha} (\bm{k},i\omega_n ; \bm{q},i\nu_m) - 1 \right] + i\sum_i q_i \left[ v_{\alpha,\bm{k}}^{(i)} \Lambda_{\alpha}^{(i)} (\bm{k},i\omega_n ; \bm{q},i\nu_m) - v_{\alpha,\bm{k}}^{(i)} \right] = \frac{\Theta_{n,m}}{\tau_{\alpha,\bm{k}}^\textrm{qp}}.
\end{equation}

Inserting $\bm{q} = \bm{0}$ into Eq.~\eqref{eq:charge_vertex_dyson_equation_imaginary_alternative_pre}, we find that the charge vertex $\Lambda_{0\alpha} (\bm{k},i\omega_n ; \bm{q},i\nu_m)$ at $\bm{q}=0$ is given by Eq.~\eqref{eq:charge_vertex_q=0}. On the other hand, inserting Eq.~\eqref{eq:charge_current_dyson_equation_sum} to \eqref{eq:ward_identity}, we obtain
\begin{equation}
	\label{eq:ward_identity_alternative}
	\mathcal{G}_{\alpha}^{-1} (\bm{k} + \bm{q},i\omega_n + i\nu_m) - \mathcal{G}_{\alpha}^{-1} (\bm{k},i\omega_n) = i\nu_m - \bm{q} \cdot \bm{v}_{\alpha,\bm{k}} + \Theta_{n,m} \frac{i}{\tau_{\alpha,\bm{k}}^\textrm{qp}},
\end{equation}
resulting in Eq.~\eqref{eq:green_function_product_imaginary_denominator} in the range where $\Theta_{n,m} = 1$. Note that Eq.~\eqref{eq:ward_identity_alternative} can be alternatively obtained from the approximate form of the self-energy given by \cite{S_Coleman2016, S_Flensberg2004}
\begin{equation}
	\Sigma_\alpha (\bm{k},i\omega_n) \approx - \frac{i}{2\tau_{\alpha, \bm{k}}^\textrm{qp}} \operatorname{sgn} (\omega_n).
\end{equation}

From Eqs.~(\ref{eq:charge_vertex_ansatz})-(\ref{eq:large_nu_m}) in the main text,
$\Lambda_{0\alpha} (\bm{k},i\omega_n ; \bm{q},i\nu_m) \approx 1 + \Theta_{n,m} [1 - i \sum_i q_i (\tau_{\alpha,\bm{k}}^{(i)} - \tau_{\alpha,\bm{k}}^\textrm{qp})] / \nu_m \tau_{\alpha,\bm{k}}^\textrm{qp}$ in $\bm{q} \rightarrow \bm{0}$ limit. Applying it to Eq.~\eqref{eq:charge_current_dyson_equation_sum}, the current vertex in $\bm{q} \rightarrow \bm{0}$ limit is given by
\begin{equation}
	\label{eq:current_vertex_q=0}
	\Lambda_{\alpha}^{(i)} (\bm{k},i\omega_n ; \bm{q} \rightarrow \bm{0},i\nu_m) \approx 1 + \Theta_{n,m} \left( \frac{\tau_{\alpha,\bm{k}}^{(i)}}{\tau_{\alpha,\bm{k}}^\textrm{qp}} - 1 \right),
\end{equation}
which is consistent with the one suggested in Ref.~\cite{S_Kim2019}.

\section{Detailed derivations of the charge vertex}
\label{sec:charge_vertex_detailed_derivation}
Inserting Eq.~\eqref{eq:charge_vertex_ansatz} to \eqref{eq:charge_vertex_dyson_equation_imaginary_alternative} and expanding the right hand side, we obtain
\begin{align}
	\label{eq:charge_vertex_dyson_equation_ansatz}
	\frac{1 - i \bm{q} \cdot \bm{l}_{\alpha,\bm{k}}}{\tau_{\alpha, \bm{k}}^\textrm{qp}} & = \sum_{\alpha^\prime} \int \frac{d^d k^\prime}{(2\pi)^d} W_{\alpha,\bm{k} ; \alpha^\prime, \bm{k^\prime}} \frac{1 + \mathcal{V}_m (\bm{q},\nu_m) \tau_{\alpha^\prime, \bm{k^\prime}}^\textrm{qp} - i\bm{q} \cdot \bm{l}_{\alpha^\prime,\bm{k^\prime}}}{1 + \nu_m \tau_{\alpha^\prime, \bm{k^\prime}}^\textrm{qp} + i \bm{q} \cdot \bm{v}_{\alpha^\prime, \bm{k^\prime}} \tau_{\alpha^\prime, \bm{k^\prime}}^\textrm{qp}} \\ & = \sum_{\alpha^\prime} \int \frac{d^d k^\prime}{(2\pi)^d} W_{\alpha,\bm{k} ; \alpha^\prime, \bm{k^\prime}} \Big\{ 1 + \left[ \mathcal{V}_m (\bm{q},\nu_m) - \nu_m \right] \tau_{\alpha^\prime, \bm{k^\prime}}^\textrm{qp} -i \bm{q} \cdot \left( \bm{l}_{\alpha^\prime,\bm{k^\prime}} + \bm{v}_{\alpha^\prime,\bm{k^\prime}} \tau_{\alpha^\prime, \bm{k^\prime}}^\textrm{qp} \right) + O^2(\bm{q},\nu_m) \Big\}. \nonumber
\end{align}
Assuming a long wavelength ($\bm{q} \cdot \bm{v}_{\alpha^\prime,\bm{k^\prime}} \tau^\textrm{qp}_{\alpha^\prime,\bm{k^\prime}} \ll 1$) and low frequency ($\nu_m \tau^\textrm{qp}_{\alpha^\prime,\bm{k^\prime}} \ll 1$) limit and comparing both sides of Eq.~\eqref{eq:charge_vertex_dyson_equation_ansatz} up to linear order in $\bm{q}$ and $\nu_m$, we obtain
\begin{subequations}
	\label{eq:charge_vertex_dyson_equation_1st_order}
 	\begin{equation}
 		\mathcal{V}_m (\bm{q},\nu_m) = \nu_m + O^2(\bm{q},\nu_m),
 	\end{equation}
 	\begin{equation}
 		\label{eq:l_integral_equation}
 		\frac{\bm{l}_{\alpha,\bm{k}}}{\tau_{\alpha, \bm{k}}^\textrm{qp}} = \sum_{\alpha^\prime} \int \frac{d^d k^\prime}{(2\pi)^d} W_{\alpha,\bm{k} ; \alpha^\prime, \bm{k^\prime}} \left( \bm{l}_{\alpha^\prime,\bm{k^\prime}} + \bm{v}_{\alpha^\prime,\bm{k^\prime}} \tau_{\alpha^\prime, \bm{k^\prime}}^\textrm{qp} \right).
 	\end{equation}
\end{subequations}
Identifying Eq.~\eqref{eq:l_integral_equation} and the integral equation for the transport relaxation time given by \cite{S_Park2017, S_Kim2019}
\begin{equation}
	\label{eq:transport_relaxation_time}
	v_{\alpha,\bm{k}}^{(i)} \left( \frac{\tau_{\alpha,\bm{k}}^{(i)}}{\tau_{\alpha,\bm{k}}^\textrm{qp}} - 1 \right) = \sum_{\alpha^\prime} \int \frac{d^d k^\prime}{(2\pi)^d} W_{\alpha,\bm{k} ; \alpha^\prime, \bm{k^\prime}} v^{(i)}_{\alpha^\prime,\bm{k^\prime}} \tau_{\alpha^\prime, \bm{k^\prime}}^{(i)},
\end{equation}
where $v_{\alpha,\bm{k}}^{(i)}$ and $\tau_{\alpha,\bm{k}}^{(i)}$ are the $i$-th component of the velocity and transport relaxation time, respectively, we obtain the $i$-th direction of $\bm{l}_{\alpha,\bm{k}}$ as follows:
\begin{equation}
	\label{eq:l_ith_component_SM}
	l^{(i)}_{\alpha,\bm{k}} = v_{\alpha,\bm{k}}^{(i)} \left( \tau_{\alpha,\bm{k}}^{(i)} - \tau_{\alpha,\bm{k}}^\textrm{qp} \right).
\end{equation}
To evaluate $\mathcal{V}_m (\bm{q},\nu_m)$, we take the average of Eq.~\eqref{eq:charge_vertex_dyson_equation_ansatz} over the Fermi surface, resulting in
\begin{equation}
	\label{eq:charge_vertex_dyson_equation_ansatz_average}
	0 = \sum_{\alpha} \int \frac{d^d k}{(2\pi)^d} \delta (\xi_{\alpha,\bm{k}}) \Big\{ \left[ \mathcal{V}_m (\bm{q},\nu_m) - \nu_m \right] - \sum_{i,j} q_i q_j v_{\alpha,\bm{k}}^{(i)} v_{\alpha,\bm{k}}^{(j)} \tau_{\alpha,\bm{k}}^{(j)} + O^3(\bm{q},\nu_m) \Big\},
\end{equation}
Here we assume $\xi_{\alpha,\bm{k}} = \xi_{\alpha,-\bm{k}}$ to cancel off the linear terms in velocities. Identifying both sides of Eq.~\eqref{eq:charge_vertex_dyson_equation_ansatz_average}, we obtain
\begin{equation}
	\label{eq:large_nu_m_SM}
	\mathcal{V}_m (\bm{q},\nu_m) = \nu_m + \sum_{i,j} q_i q_j \mathcal{D}_{ij}  + O^3(\bm{q},\nu_m).
\end{equation}
Here, $\mathcal{D}_{ij}$ is the diffusion constant defined by Eq.~\eqref{eq:diffusion_constant} in the main text.

\section{Alternative derivations for the vertex corrections}
\label{sec:alternative_derivation}
In this section, we present an alternative diagrammatic approach performing the frequency summation first to obtain the density-density response function and corresponding diffusion constant for $d$-dimensional anisotropic multiband systems, motivated from the method depicted in Ref.~\cite{S_Kim2019}.

\begin{figure}[htb]
	\includegraphics[width=.5\linewidth]{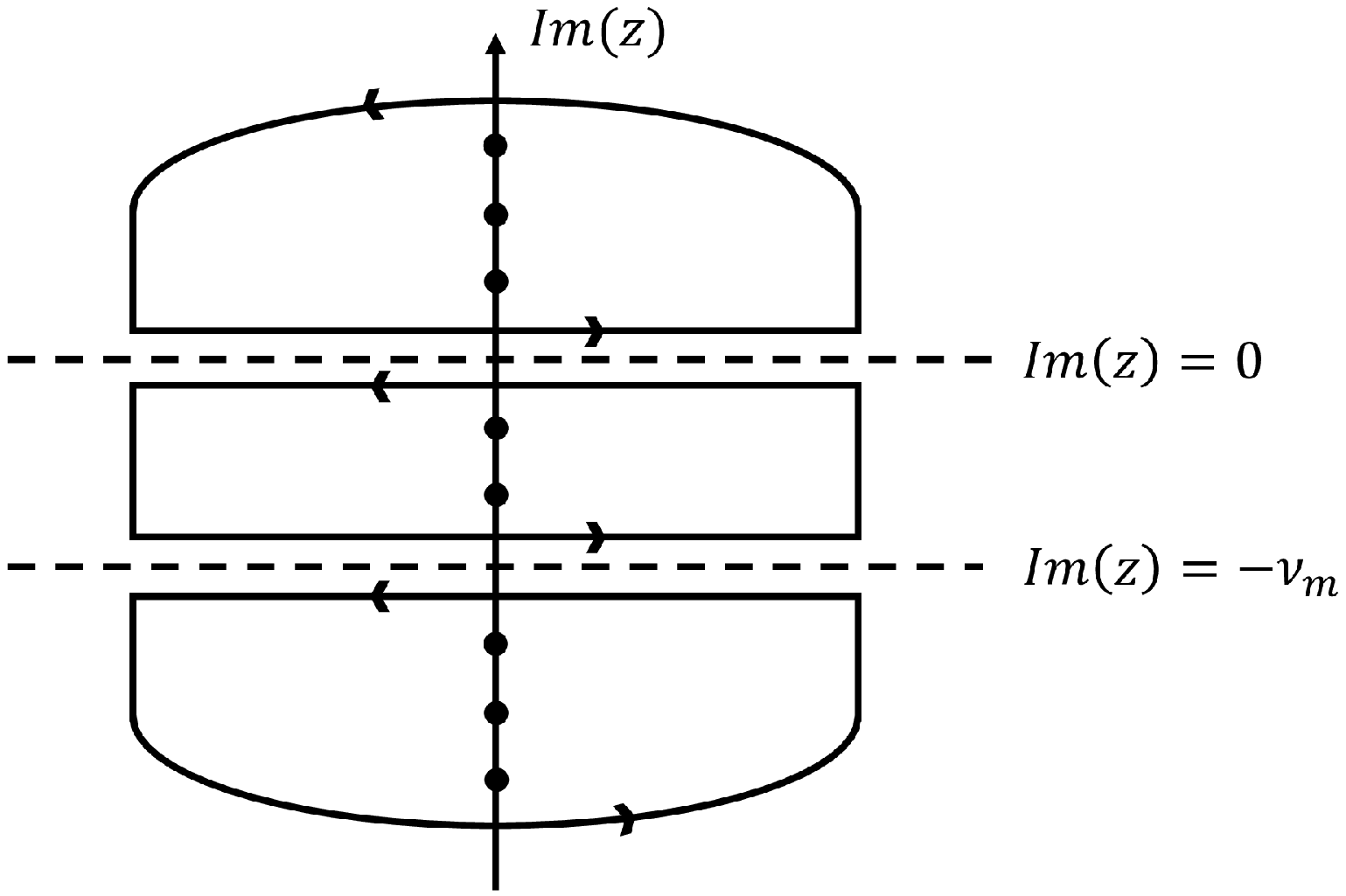}
	\caption{Contour $C$ for the integration described in Eq.~\eqref{eq:density_response_contour}. Note that there are two branch cuts along the dashed line and the dots on the $\mathrm{Im} (z)$-axis represent the poles at $z = i\omega_n$.} 
	\label{fig:contour_C}
\end{figure}

The contribution of the vertex corrections to the density response given by Eq.~\eqref{eq:density_response_1_vertex_imaginary} can be written as the following contour integral form:
\begin{equation}
	\label{eq:density_response_contour}
	\chi_1(\bm{q},i\nu_m) = \frac{1}{\beta} \sum_{i\omega_n} P(\bm{q}; i\omega_n, i\omega_n + i\nu_m) = -\oint_C \frac{dz}{2\pi i} f^{(0)} (z) P(\bm{q};z,z+i\nu_m),
\end{equation}
where $f^{(0)} (z) \equiv (e^{\beta z} + 1)^{-1}$ is the Fermi-Dirac distribution function and the contour $C$ is illustrated in Fig.~\ref{fig:contour_C}. For complex numbers $z$ and $w$, $P(\bm{q};z,z+w)$ is defined by
\begin{equation}
	\label{eq:P_definition}
	P(\bm{q}; z, z+w) \equiv \mathrm{g} \sum_\alpha \int \frac{d^d k}{(2\pi)^d} \left[ \Lambda_{0 \alpha} (\bm{k},z ; \bm{q},w) -1 \right] \mathcal{G}_\alpha (\bm{k},z) \mathcal{G}_\alpha (\bm{k} + \bm{q} , z + w).
\end{equation}
Specifying the integral along the contour $C$, Eq.~\eqref{eq:density_response_contour} is rewritten as
\begin{align}
	\label{eq:density_response_contour_alternative}
	\chi_1 (\bm{q},i\nu_m) = \int \frac{d\xi}{2\pi i} f^{(0)} (\xi) \Big[ &- P(\bm{q}; \xi + i0^+ , \xi + i\nu_m) + P(\bm{q}; \xi - i0^- , \xi + i\nu_m) \\ & - P(\bm{q}; \xi - i\nu_m , \xi + i0^+) + P(\bm{q}; \xi - i\nu_m , \xi - i0^+) \Big]. \nonumber
\end{align}
Taking the analytic continuation $i\nu_m \rightarrow \nu + i0^+$ and assuming the low frequency limit, Eq.~\eqref{eq:density_response_contour_alternative} transforms into
\begin{equation}
	\label{eq:retarded_density_response_P}
	\chi_1^\textrm{R} (\bm{q},\nu) = \int \frac{d\xi}{2\pi i} \left[ \nu S^{(0)} (\xi) P^\textrm{AR} (\bm{q}; \xi,\xi+\nu) - f^{(0)} (\xi) P^\textrm{RR} (\bm{q}; \xi,\xi+\nu) + f^{(0)} (\xi + \nu) P^{\textrm{RR} \star} (\bm{q}; \xi,\xi+\nu) \right],
\end{equation}
where $S^{(0)} \equiv - \frac{df^{(0)} (\xi)}{d\xi}$, the superscripts $\textrm{A}$ and $\textrm{R}$ represent the advanced and retarded functions, respectively, and $P^\textrm{AR(RR)}$ is given as follows by taking the analytic continuation to Eq.~\eqref{eq:P_definition}:
\begin{equation}
	\label{eq:P_AR_RR}
	P^\textrm{AR(RR)} (\bm{q}; \xi, \xi + \nu) = \mathrm{g} \sum_\alpha \int \frac{d^d k}{(2\pi)^d} \left[ \Lambda_{0 \alpha}^\textrm{AR(RR)} (\bm{k},\xi ; \bm{q}, \nu) - 1 \right] \mathcal{G}_\alpha^\textrm{A(R)} (\bm{k},\xi) \mathcal{G}_\alpha^\textrm{R} (\bm{k} + \bm{q}, \xi + \nu).
\end{equation}
Note that $P^{\textrm{AA}} = P^{\textrm{RR}\star}$ is used, where $\star$ represents the complex conjugation. On the other hand, taking the analytic continuation $i\omega_n \rightarrow \xi \mp i0^+$ and $i\omega_n + i\nu_m \rightarrow \xi + \nu + i0^+$ to Eq.~\eqref{eq:charge_vertex_dyson_equation_imaginary}, the Dyson equation for the vertex correction is given by
\begin{equation}
	\label{eq:charge_vertex_dyson_equation}
	\Lambda_{0 \alpha}^\textrm{AR(RR)} (\bm{k},\xi;\bm{q},\nu) - 1  = n_\textrm{imp} \sum_{\alpha^\prime} \int \frac{d^d k^\prime}{(2\pi)^d} \left| V_{\alpha,\bm{k};\alpha^\prime,\bm{k^\prime}} \right|^2 \Lambda_{0 \alpha^\prime}^\textrm{AR(RR)} (\bm{k^\prime},\xi ; \bm{q},\nu) \mathcal{G}_{\alpha^\prime}^\textrm{A(R)} (\bm{k^\prime},\xi) \mathcal{G}_{\alpha^\prime}^\textrm{R} (\bm{k^\prime} + \bm{q}, \xi + \nu).
\end{equation}
Since $\mathcal{G}_\alpha^\textrm{R} (\bm{k},\xi) \mathcal{G}_\alpha^\textrm{R} (\bm{k},\xi)$ vanishes in the low impurity density limit, the contribution of $P^\textrm{RR}$ becomes negligible in the low frequency-long wavelength limit \cite{S_Kim2019, S_Mahan2000}. On the other hand, to evaluate the contribution of $P^\textrm{AR}$, let us begin with rewriting $\mathcal{G}_\alpha^\textrm{R} (\bm{k} + \bm{q}, \xi + \nu)$ as
\begin{align}
	\label{eq:retarded_green_function_deviation}
	\mathcal{G}_\alpha^\textrm{R} (\bm{k} + \bm{q}, \xi + \nu) & = \mathcal{G}_\alpha^\textrm{R} (\bm{k}, \xi) \left[ 1 + \eta_{\alpha,\bm{k}} (\xi;\bm{q},\nu) \mathcal{G}_\alpha^\textrm{R} (\bm{k}, \xi) \right]^{-1} \nonumber \\ & = \mathcal{G}_\alpha^\textrm{R} (\bm{k}, \xi) \Big\{ 1 - \eta_{\alpha,\bm{k}} (\xi;\bm{q},\nu) \mathcal{G}_\alpha^\textrm{R} (\bm{k}, \xi) + \left[ \eta_{\alpha,\bm{k}} (\xi;\bm{q},\nu) \mathcal{G}_\alpha^\textrm{R} (\bm{k}, \xi) \right]^2 + \cdots \Big\},
\end{align}
where $\eta_{\alpha,\bm{k}} (\xi ; \bm{q},\nu)$ is defined by
\begin{equation}
	\label{eq:eta_definition}
	\eta_{\alpha,\bm{k}} (\xi ; \bm{q},\nu) = \mathcal{G}_\alpha^\textrm{R} (\bm{k} + \bm{q}, \xi + \nu)^{-1} - \mathcal{G}_\alpha^\textrm{R} (\bm{k}, \xi)^{-1}.
\end{equation}
Then, $\mathcal{G}_\alpha^\textrm{A} (\bm{k},\xi) \mathcal{G}_\alpha^\textrm{R} (\bm{k} + \bm{q},\xi + \nu)$ is given as follows in the low impurity density limit in which the self-energy is negligibly small:
\begin{align}
	\label{eq:green_function_product_AR}
	\mathcal{G}_\alpha^\textrm{A} (\bm{k},\xi) \mathcal{G}_\alpha^\textrm{R} (\bm{k} + \bm{q},\xi + \nu) & = \frac{1}{\omega_{\alpha,\bm{k}}^2 + \Delta_{\alpha,\bm{k}}^2 (\xi)} \Bigg\{ 1 - \frac{\eta_{\alpha,\bm{k}} (\xi; \bm{q}, \nu)}{\omega_{\alpha,\bm{k}} + i\Delta_{\alpha,\bm{k}} (\xi)} + \frac{\eta_{\alpha,\bm{k}}^2 (\xi; \bm{q}, \nu)}{\left[ \omega_{\alpha,\bm{k}} + i\Delta_{\alpha,\bm{k}} (\xi) \right]^2 } + \cdots \Bigg\} \nonumber \\ & \approx \frac{\pi}{\Delta_{\alpha,\bm{k}}} \delta (\omega_{\alpha,\bm{k}}) \left[ 1 + \frac{i\eta_{\alpha,\bm{k}} (\bm{q}, \nu)}{2\Delta_{\alpha,\bm{k}}} - \frac{\eta_{\alpha,\bm{k}}^2 (\bm{q}, \nu)}{4\Delta_{\alpha,\bm{k}}^2} +\cdots \right], 
\end{align}
where $\omega_{\alpha,\bm{k}} \equiv \xi - \xi_{\alpha,\bm{k}}$, $\Delta_{\alpha,\bm{k}} (\xi) \equiv \operatorname{Im} \Sigma_\alpha^\textrm{A} (\bm{k}, \xi)$, $\Delta_{\alpha,\bm{k}} \equiv \Delta_{\alpha,\bm{k}} (\xi_{\alpha,\bm{k}})$, and $\eta_{\alpha,\bm{k}} (\bm{q}, \nu)\equiv \eta_{\alpha,\bm{k}} (\xi_{\alpha,\bm{k}}; \bm{q}, \nu)$. Here, the real part of the self-energy is integrated into the definition of the chemical potential \cite{S_Brouwer2005, S_Flensberg2004}. From $\operatorname{Im} \Sigma_\alpha^\textrm{R} (\bm{k},\xi_{\alpha,\bm{k}}) \approx - 1 / 2\tau_{\alpha, \bm{k}}^\textrm{qp}$ up to the Born approximation \cite{S_Flensberg2004, S_Brouwer2005}, we obtain $\Delta_{\alpha,\bm{k}} \approx 1 / 2\tau_{\alpha,\bm{k}}^\textrm{qp}$. Furthermore, the low impurity density limit leads to $|\bm{v}_{\alpha,\bm{k}}| \gg |\frac{\partial}{\partial \bm{k}} \Delta_{\alpha, \bm{k}}|$, resulting in
\begin{equation}
	\label{eq:eta_evaluation}
	\eta_{\alpha,\bm{k}} (\bm{q},\nu) \approx \nu - \bm{q} \cdot \bm{v}_{\alpha,\bm{k}}.
\end{equation}
Inserting Eq.~\eqref{eq:green_function_product_AR} into Eq.~\eqref{eq:P_AR_RR}, $P^\textrm{AR}$ is given by
\begin{align}
	\label{eq:P_AR}
	P^\textrm{AR} (\bm{q};\xi, \xi + \nu) & = 2\pi \mathrm{g} \sum_\alpha \int \frac{d^d k}{(2\pi)^d} \delta(\xi - \xi_{\alpha,\bm{k}}) \tau_{\alpha,\bm{k}}^\textrm{qp} \left[ \Lambda_{0 \alpha}^\textrm{AR} (\bm{k},\xi_{\alpha,\bm{k}} ; \bm{q},\nu) - 1 \right] \nonumber \\ & \times \left\{ 1 + i (\nu - \bm{q} \cdot \bm{v}_{\alpha,\bm{k}}) \tau_{\alpha,\bm{k}}^\textrm{qp} - \left[ (\nu - \bm{q} \cdot \bm{v}_{\alpha,\bm{k}} ) \tau_{\alpha,\bm{k}}^\textrm{qp} \right]^2 + O^3 (\bm{q},\nu) \right\},
\end{align}
so that Eq.~\eqref{eq:retarded_density_response_P} transforms into
\begin{align}
	\label{eq:retarded_density_response_expansion}
	\chi_1^\textrm{R} (\bm{q},\nu) & = - i \nu \sum_\alpha \int \frac{d^d k}{(2\pi)^d} \delta(\xi_{\alpha,\bm{k}}) \tau_{\alpha, \bm{k}}^\textrm{qp} \left[ \Lambda_{0 \alpha}^\textrm{AR} (\bm{k},\xi_{\alpha, \bm{k}} ; \bm{q},\nu) - 1 \right] \nonumber \\ & \times \left\{ 1 + i (\nu - \bm{q} \cdot \bm{v}_{\alpha,\bm{k}}) \tau_{\alpha,\bm{k}}^\textrm{qp} - \left[ (\nu - \bm{q} \cdot \bm{v}_{\alpha,\bm{k}} ) \tau_{\alpha,\bm{k}}^\textrm{qp} \right]^2 + O^3 (\bm{q},\nu) \right\}.
\end{align}
Here, we adopt the low temperature approximation $S^{(0)} (\xi) \approx \delta(\xi)$. To evaluate $\Lambda_{0 \alpha}^\textrm{AR} (\bm{k},\xi_{\alpha, \bm{k}} ; \bm{q},\nu)$, we insert Eq.~\eqref{eq:green_function_product_AR} into Eq.~\eqref{eq:charge_vertex_dyson_equation} with $\xi = \xi_{\alpha, \bm{k}}$, resulting in
\begin{align}
	\label{eq:charge_vertex_dyson_equation_expansion}
	\Lambda_{0 \alpha}^\textrm{AR} (\bm{k},\xi_{\alpha, \bm{k}};\bm{q},\nu) - 1 & = \sum_{\alpha^\prime} \int \frac{d^d k^\prime}{(2\pi)^d}  W_{\alpha,\bm{k};\alpha^\prime,\bm{k^\prime}} \tau_{\alpha^\prime, \bm{k^\prime}}^\textrm{qp} \Lambda_{0 \alpha^\prime}^\textrm{AR} (\bm{k^\prime},\xi_{\alpha^\prime, \bm{k^\prime}};\bm{q},\nu) \nonumber \\ & \times \bigg\{ 1 + i (\nu - \bm{q} \cdot \bm{v}_{\alpha^\prime,\bm{k^\prime}}) \tau_{\alpha^\prime,\bm{k^\prime}}^\textrm{qp} - \left[ (\nu - \bm{q} \cdot \bm{v}_{\alpha^\prime,\bm{k^\prime}} ) \tau_{\alpha^\prime,\bm{k^\prime}}^\textrm{qp} \right]^2 + O^3 (\bm{q},\nu) \bigg\}.
\end{align}
Inserting $\bm{q} = \bm{0}$, Eq.~\eqref{eq:charge_vertex_dyson_equation_expansion} transforms into
\begin{equation}
	\label{eq:charge_vertex_dyson_equation_expansion_q=0}
	\Lambda_{0 \alpha}^\textrm{AR} (\bm{k},\xi_{\alpha, \bm{k}};\bm{0},\nu) - 1 = \frac{i}{\nu} \sum_{\alpha^\prime} \int \frac{d^d k^\prime}{(2\pi)^d}  W_{\alpha,\bm{k};\alpha^\prime,\bm{k^\prime}} \Lambda_{0 \alpha^\prime}^\textrm{AR} (\bm{k^\prime},\xi_{\alpha^\prime, \bm{k^\prime}};\bm{0},\nu) \left( 1 + \frac{i}{\nu \tau_{\alpha^\prime,\bm{k^\prime}}^\textrm{qp}} \right)^{-1}.
\end{equation}
Here, we use $1+x+x^2+O^3(x) \approx (1-x)^{-1}$. Using Eq.~\eqref{eq:quasiparticle_lifetime}, we can infer that
\begin{equation}
	\label{eq:charge_vertex_AR_q=0}
	\Lambda_{0 \alpha}^\textrm{AR} (\bm{k},\xi_{\alpha, \bm{k}};\bm{0},\nu) = 1 + \frac{i}{\nu \tau_{\alpha,\bm{k}}^\textrm{qp}}.
\end{equation}
Similarly as in Eq.~(\ref{eq:charge_vertex_ansatz}) in the main text, we assume the following ansatz for $\Lambda_{0 \alpha}^\textrm{AR} (\bm{k},\xi_{\alpha, \bm{k}};\bm{0},\nu)$:
\begin{equation}
	\label{eq:charge_vertex_AR_ansatz}
	\Lambda_{0 \alpha}^\textrm{AR} (\bm{k},\xi_{\alpha, \bm{k}};\bm{q},\nu) = 1 + \frac{i(1-i\bm{q} \cdot \bm{l}_{\alpha,\bm{k}})}{\mathcal{V} (\bm{q},\nu) \tau_{\alpha,\bm{k}}^\textrm{qp}}
\end{equation}
for some $\bm{l}_{\alpha,\bm{k}}$ and $\mathcal{V} (\bm{q},\nu)$ satisfying $\mathcal{V} (\bm{0},\nu) = \nu$. Inserting Eq.~\eqref{eq:charge_vertex_AR_ansatz} into \eqref{eq:charge_vertex_dyson_equation_expansion}, we obtain
\begin{equation}
	\label{eq:charge_vertex_AR_dyson_equation_ansatz}
	\frac{1 - i\bm{q} \cdot \bm{l}_{\alpha,\bm{k}}}{\tau_{\alpha,\bm{k}}^\textrm{qp}} = \sum_{\alpha^\prime} \int \frac{d^d k^\prime}{(2\pi)^d}  W_{\alpha,\bm{k};\alpha^\prime,\bm{k^\prime}} \Big\{ 1 + i \left[ \nu - \mathcal{V}(\bm{q},\nu) \right] \tau_{\alpha^\prime, \bm{k^\prime}}^\textrm{qp} - i\bm{q} \cdot (\bm{l}_{\alpha^\prime, \bm{k^\prime}} + \bm{v}_{\alpha^\prime, \bm{k^\prime}} \tau_{\alpha^\prime, \bm{k^\prime}}^\textrm{qp}) + O^2 (\bm{q},\nu) \Big\}.
\end{equation}
Identifying the both sides of Eq.~\eqref{eq:charge_vertex_AR_dyson_equation_ansatz} up to linear order in $\bm{q}$ and $\nu_m$, we obtain $\mathcal{V} (\bm{q},\nu) = \nu + O^2(\bm{q},\nu)$ and Eq.~\eqref{eq:l_integral_equation}, which results in $\bm{l}_{\alpha,\bm{k}}$ given by Eq.~\eqref{eq:l_ith_component}. On the other hand, inserting Eq.~\eqref{eq:charge_vertex_AR_ansatz} into \eqref{eq:charge_vertex_dyson_equation_expansion} averaged over the surface of energy $\xi_{\alpha, \bm{k}}$, we obtain
\begin{equation}
	\label{eq:charge_vertex_AR_dyson_equation_ansatz_averaged}
	0 = \sum_{\alpha} \int \frac{d^d k}{(2\pi)^d} \delta (\xi_{\alpha,\bm{k}}) \Big\{ i \left[ \nu - \mathcal{V} (\bm{q},\nu) \right] - \sum_{i,j} q_i q_j v_{\alpha,\bm{k}}^{(i)} v_{\alpha,\bm{k}}^{(j)} \tau_{\alpha,\bm{k}}^{(j)} + O^3(\bm{q},\nu) \Big\},
\end{equation}
resulting in
\begin{equation}
	\label{eq:large_nu}
	\mathcal{V} (\bm{q},\nu) = \nu + i \sum_{i,j} q_i q_j \mathcal{D}_{ij} + O^3(\bm{q},\nu),
\end{equation}
where $\mathcal{D}_{ij}$ is the diffusion constant given by Eq.~\eqref{eq:diffusion_constant}. Inserting Eq.~\eqref{eq:charge_vertex_AR_ansatz} to \eqref{eq:retarded_density_response_expansion}, $\chi_1^\textrm{R} (\bm{q},\nu)$ finally reduces to
\begin{equation}
	\label{eq:retarded_density_response_1_final}
	\chi_1^\textrm{R} (\bm{q},\nu) = N(0) \frac{\nu \left[ 1 + O^1 (\bm{q},\nu) \right]}{\nu + i\sum_{i,j} q_i q_j \mathcal{D}_{ij} + O^3(\bm{q},\nu)}.
\end{equation}
Therefore, up to leading order in $\bm{q}$ and $\nu$, $\chi^\textrm{R} (\bm{q},\nu) = \chi_0^\textrm{R} (\bm{q},\nu) + \chi_1^\textrm{R} (\bm{q},\nu)$ is given by
\begin{equation}
	\chi^{(\textrm{R})} (\bm{q},\nu) = N(0) \frac{ i\sum_{i,j} q_i q_j \mathcal{D}_{ij}}{\nu + i \sum_{i,j} q_i q_j \mathcal{D}_{ij}},
\end{equation}
which is consistent with Eq.~\eqref{eq:density_response_retarded}.

\section{Calculations for the diffusion constants}
\label{sec:diffusion_constants_calculation}
\subsection{Anisotropic 2D electron gas}
\label{sec:diffusion_constants_calculation_electron_gas}
For anisotropic 2D electron gas (2DEG), the Hamiltonian is given by
\begin{equation}
	H = \frac{k_x^2}{2m_x} + \frac{k_y^2}{2m_y},
\end{equation}
where $m_x$ and $m_y$ are the effective masses along the $x$ and $y$ directions, respectively. Under the coordinate transformation $(k_x , k_y) = (\sqrt{2m_x \varepsilon}\cos\theta , \sqrt{2m_y \varepsilon} \sin\theta)$ with the Jacobian $\mathcal{J} (\varepsilon,\theta) = \sqrt{m_x m_y}$, the Hamiltonian becomes $H = \varepsilon$, and the $x$ and $y$ components of the velocity are given by $v^{(x)} (\varepsilon,\theta) = \sqrt{2\varepsilon/m_x} \cos\theta$ and $v^{(y)} (\varepsilon,\theta) = \sqrt{2\varepsilon/m_y} \sin\theta$, respectively. Note that the density of states is given by
\begin{equation}
	N(\varepsilon) = \mathrm{g} \frac{\sqrt{m_x m_y}}{2\pi},
\end{equation}
where $\mathrm{g} = 2$ is the spin degeneracy factor.

The short-range impurity potential is given by $V(\bm{q}) = V_0$ in the momentum space, where $\bm{q} \equiv \bm{k} - \bm{k}^\prime$. Inserting the transition rate $W_{\bm{k};\bm{k}^\prime} = 2\pi n_\textrm{imp} \left| V(\bm{q}) \right|^2 \delta(\varepsilon-\varepsilon^\prime)$ into Eq.~\eqref{eq:quasiparticle_lifetime}, the quasiparticle lifetime is given by
\begin{equation}
	\frac{1}{\tau^\textrm{qp}} = n_\textrm{imp} V_0^2 \sqrt{m_x m_y}.
\end{equation}
On the other hand, inserting $W_{\bm{k};\bm{k}^\prime}$ into Eq.~\eqref{eq:transport_relaxation_time}, we obtain the following integral equations for the transport relaxation time:
\begin{subequations}
	\begin{equation}
		\label{eq:electron_gas_tau_x}
		\left[ \tau^{(x)} (\varepsilon, \theta) - \tau^\textrm{qp} \right] \cos\theta = \int_0^{2\pi} \frac{d\theta^\prime}{2\pi} \tau^{(x)} (\varepsilon, \theta^\prime) \cos\theta^\prime,
	\end{equation}
	\begin{equation}
		\label{eq:electron_gas_tau_y}
		\left[ \tau^{(y)} (\varepsilon, \theta) - \tau^\textrm{qp} \right] \sin\theta = \int_0^{2\pi} \frac{d\theta^\prime}{2\pi} \tau^{(y)} (\varepsilon, \theta^\prime) \sin\theta^\prime.
	\end{equation}
\end{subequations}
The right-hand sides of Eqs.~\eqref{eq:electron_gas_tau_x} and \eqref{eq:electron_gas_tau_y}, which are proportional to the average values of $\tau^{(x)} v^{(x)}$ and $\tau^{(y)} v^{(y)}$, respectively, vanish owing to the assumption $\varepsilon(\bm{k}) = \varepsilon(-\bm{k})$. Hence, we have
\begin{equation}
	\tau^{(x)} = \tau^{(y)} = \tau^\textrm{qp}.
\end{equation}
Thus, from Eqs.~\eqref{eq:diffusion_constant}, we obtain
\begin{subequations}
	\begin{equation}
		\mathcal{D}_{xx} = \frac{\varepsilon_\textrm{F}}{m_x} \tau^\textrm{qp},
	\end{equation}
	\begin{equation}
		\mathcal{D}_{yy} = \frac{\varepsilon_\textrm{F}}{m_y} \tau^\textrm{qp},
	\end{equation}
\end{subequations}
where $\varepsilon_\textrm{F}$ is the Fermi energy.

On the other hand, the long-range Coulomb impurity potential within the Thomas-Fermi approximation is given by
\begin{equation}
	\label{eq:screened_Coulomb_impurity_potential}
	V(\bm{q}) = \frac{2\pi e^2}{\epsilon_0 (q + q_\textrm{TF})},
\end{equation}
where $q \equiv |\bm{q}|$, $\epsilon_0$ is the background dielectric constant, and $q_\textrm{TF}$ is the Thomas-Fermi wavevector given by
\begin{equation}
	q_\textrm{TF} = \frac{2\pi e^2}{\epsilon_0} N(\varepsilon_\textrm{F}) = \frac{2e^2}{\epsilon_0} \sqrt{m_x m_y}.
\end{equation}
Inserting $W_{\bm{k};\bm{k}^\prime} = 2\pi n_\textrm{imp} \left| V(\bm{q}) \right|^2 \delta(\varepsilon-\varepsilon^\prime)$ into Eq.~\eqref{eq:quasiparticle_lifetime}, the quasiparticle lifetime at the Fermi energy is given by
\begin{equation}
	\label{eq:electron_gas_tau_qp_long}
	\frac{\tau_0}{\tau^\textrm{qp} (\varepsilon_\textrm{F},\theta)} = \int_0^{2\pi} d\theta^\prime \frac{1}{\left[ \tilde{q} (\theta, \theta^\prime) + Q \right]^2}.
\end{equation}
Here, $\tau_0 \equiv \varepsilon_\textrm{F} / \varepsilon_0^2$, $Q \equiv q_\textrm{TF}/k_\textrm{F}$, and $\tilde{q}(\theta,\theta^\prime)$ is defined by
\begin{equation}
	\tilde{q}(\theta,\theta^\prime) \equiv \left[ A (\cos\theta - \cos\theta^\prime)^2 + \frac{1}{A} (\sin\theta - \sin\theta^\prime)^2  \right]^{\frac{1}{2}},
\end{equation}
where $\varepsilon_0 \equiv \frac{e^2}{\epsilon_0 d_0}$, $d_0 \equiv (\pi n_\textrm{imp})^{-\frac{1}{2}}$ is the average distance between impurities, $A \equiv k_{\textrm{F}}^{(x)} / k_{\textrm{F}}^{(y)} = \sqrt{m_x / m_y}$ is the anisotropy factor, $k_{\textrm{F}}^{(i)}$ is the Fermi wavevector along the $i$-th direction, and $k_\textrm{F}$ is the effective Fermi wavevector defined by mapping $\pi k_\textrm{F}^2$ to the area inside the Fermi surface so that
\begin{equation}
	\pi k_\textrm{F}^2 = 2\pi \varepsilon_\textrm{F} \sqrt{m_x m_y}.
\end{equation}
On the other hand, inserting $W_{\bm{k};\bm{k}^\prime}$ into Eq.~\eqref{eq:transport_relaxation_time}, we obtain
\begin{subequations}
	\begin{align}
		\label{eq:electron_gas_tau_x_long}
		\left[ \frac{\tau^{(x)} (\varepsilon_\textrm{F},\theta)}{\tau^\textrm{qp} (\varepsilon_\textrm{F},\theta)} - 1 \right] \cos\theta = \frac{1}{\tau_0} \int_0^{2\pi} d\theta^\prime \frac{\tau^{(x)} (\varepsilon_\textrm{F},\theta^\prime) \cos\theta^\prime}{\left[ \tilde{q} (\theta, \theta^\prime) + Q \right]^2},
	\end{align}
	\begin{align}
		\label{eq:electron_gas_tau_y_long}
		\left[ \frac{\tau^{(y)} (\varepsilon_\textrm{F},\theta)}{\tau^\textrm{qp} (\varepsilon_\textrm{F},\theta)} - 1 \right] \sin\theta = \frac{1}{\tau_0} \int_0^{2\pi} d\theta^\prime \frac{\tau^{(y)} (\varepsilon_\textrm{F},\theta^\prime) \sin\theta^\prime}{\left[ \tilde{q} (\theta, \theta^\prime) + Q \right]^2} .
	\end{align}
\end{subequations}
Now, we write the transport relaxation time as
\begin{equation}
	\frac{\tau^{(i)} (\varepsilon_\textrm{F}, \theta)}{\tau_0} = \sum_n a_n^{(i)} \cos(2n\theta),
\end{equation}
where $i = x,y$ and $n=0,1,2,\cdots$. Note that only $2n$ factors are possible by the assumption $\varepsilon(\bm{k}) = \varepsilon(-\bm{k})$. Inserting a distinct set of angles $\{ \theta_m \}$ into Eqs.~\eqref{eq:electron_gas_tau_x_long} and \eqref{eq:electron_gas_tau_y_long}, we obtain the following linear equations:
\begin{subequations}
	\label{eq:electron_gas_relaxation_time_matrix_equation}
	\begin{align}
		\sum_n M_{mn}^{(x)} a_n^{(x)} & = \cos\theta_m,
	\end{align}
	\begin{align}
		\sum_n M_{mn}^{(y)} a_n^{(y)} & = \sin\theta_m,
	\end{align}
\end{subequations}
where the matrix elements $M_{mn}^{(x,y)}$ are given by
\begin{subequations}
	\begin{align}
		M_{mn}^{(x)} = \frac{\cos(2n\theta_m)\cos\theta_m}{\tau^\textrm{qp} (\varepsilon_\textrm{F}, \theta_m) / \tau_0} - \int_0^{2\pi} d\theta^\prime \frac{\cos\theta^\prime \cos(2n\theta^\prime)}{\left[ \tilde{q} (\theta_m, \theta^\prime) + Q \right]^2},
	\end{align}
	\begin{align}
		M_{mn}^{(y)} = \frac{\cos(2n\theta_m)\sin\theta_m}{\tau^\textrm{qp} (\varepsilon_\textrm{F}, \theta_m) / \tau_0} - \int_0^{2\pi} d\theta^\prime \frac{\sin\theta^\prime \cos(2n\theta^\prime)}{\left[ \tilde{q} (\theta_m, \theta^\prime) + Q \right]^2}.
	\end{align}
\end{subequations}
Hence, we can obtain $a_n^{(i)}$ by solving Eq.~\eqref{eq:electron_gas_relaxation_time_matrix_equation} with a large enough cutoff for $m$ and $n$. Finally, from Eq.~\eqref{eq:diffusion_constant}, we have
\begin{subequations}
	\label{eq:diffusion_constant_2DEG_numerical}
	\begin{equation}
		\mathcal{D}_{xx} = \frac{\mathcal{D}_0}{2\pi A} \int_0^{2\pi} d\theta \cos^2 \theta \left[ \sum_n a_n^{(x)} \cos(2n\theta) \right],
	\end{equation}
	\begin{equation}
		\mathcal{D}_{yy} = \frac{\mathcal{D}_0 A}{2\pi} \int_0^{2\pi} d\theta \sin^2 \theta \left[ \sum_n a_n^{(y)} \cos(2n\theta) \right],
	\end{equation}
\end{subequations}
where $\mathcal{D}_0 \equiv \frac{2\varepsilon_\textrm{F} \tau_0}{\sqrt{m_x m_y}}$.

\subsection{Anisotropic graphene}
\label{sec:diffusion_constants_calculation_graphene}
For anisotropic graphene near the Dirac point, the Hamiltonian is given by
\begin{equation}
	H = v_x k_x \sigma_x + v_y k_y \sigma_y,
\end{equation}
where $v_x$ and $v_y$ are the band velocities along the $x$ and $y$ directions, respectively, and $\bm{\sigma}$ is a vector of Pauli matrices. Under the coordinate transformation $(k_x , k_y) = (\frac{ \varepsilon}{v_x}\cos\theta , \frac{ \varepsilon}{v_y} \sin\theta)$ with the Jacobian $\mathcal{J} ( \varepsilon,\theta) = \frac{ \varepsilon}{v_x v_y}$, the Hamiltonian becomes $H = \varepsilon (\cos\theta\sigma_x + \sin\theta\sigma_y)$ and its eigenvalues and eigenvectors are given by $E (\varepsilon,\theta) = \varepsilon$ and $\big| \varepsilon,\theta \big> = \frac{1}{\sqrt{2}} ( 1, e^{i\theta})^{\rm{t}}$ so that the overlap factor is given by $F_{\bm{k}\bm{k}^\prime} = \frac{1}{2} \left[ 1 + \cos(\theta - \theta^\prime) \right]$. Here, we assume $\varepsilon > 0$. Also, the $x$ and $y$ components of the velocity are given by $v^{(x)} (\varepsilon,\theta) = v_x \cos\theta$ and $v^{(y)} (\varepsilon,\theta) = v_y \sin\theta$, respectively, and the density of states is given by
\begin{equation}
	N(\varepsilon) = \frac{\mathrm{g} \varepsilon}{2\pi v_x v_y},
\end{equation}
where $\mathrm{g} = 4$ is the spin/valley degeneracy factor.

The short-range impurity potential is given by $V(\bm{q}) = V_0$ in the momentum space, where $\bm{q} \equiv \bm{k} - \bm{k}^\prime$. Inserting the transition rate $W_{\bm{k};\bm{k}^\prime} = 2\pi n_\textrm{imp} \left| V(\bm{q}) \right|^2 F_{\bm{k}\bm{k}^\prime} \delta(\varepsilon-\varepsilon^\prime)$ into Eq.~\eqref{eq:quasiparticle_lifetime}, the quasiparticle lifetime at the Fermi energy $\varepsilon_\textrm{F}$ is given by
\begin{equation}
	\frac{1}{\tau^\textrm{qp} (\varepsilon_\textrm{F})} = \frac{n_\textrm{imp} V_0^2 \varepsilon_\textrm{F}}{2 v_x v_y} \equiv \frac{1}{\tau^\textrm{qp}_\textrm{F}}.
\end{equation}
On the other hand, inserting $W_{\bm{k};\bm{k}^\prime}$ into Eq.~\eqref{eq:transport_relaxation_time}, we obtain the following integral equations for the transport relaxation time:
\begin{subequations}
	\begin{equation}
		\label{eq:graphene_tau_x}
		\left[ \tau^{(x)} (\varepsilon_\textrm{F},\theta) - \tau^\textrm{qp}_\textrm{F} \right] \cos\theta = \frac{1}{2\pi} \int_0^{2\pi} d\theta^\prime \left\{ \cos\theta^\prime + \cos\theta^\prime \cos(\theta - \theta^\prime) \right\} \tau^{(x)} (\varepsilon_\textrm{F}, \theta^\prime),
	\end{equation}
	\begin{equation}
		\label{eq:graphene_tau_y}
		\left[ \tau^{(y)} (\varepsilon_\textrm{F},\theta) - \tau^\textrm{qp}_\textrm{F} \right] \sin\theta = \frac{1}{2\pi} \int_0^{2\pi} d\theta^\prime \left\{ \sin\theta^\prime + \sin\theta^\prime \cos(\theta - \theta^\prime) \right\} \tau^{(y)} (\varepsilon_\textrm{F}, \theta^\prime).
	\end{equation}
\end{subequations}
From the assumption $\varepsilon(\bm{k}) = \varepsilon(-\bm{k})$, the transport relaxation times can be given by the summation series of $\cos(2n\theta)$. Since the right-hand sides of Eqs.~\eqref{eq:graphene_tau_x} and \eqref{eq:graphene_tau_y} have the period $2\pi$ for $\theta$, all except $n = 0$ vanish so that we have
\begin{equation}
	\tau^{(x)} (\varepsilon_\textrm{F}) = \tau^{(y)} (\varepsilon_\textrm{F}) = 2\tau^\textrm{qp}_\textrm{F}.
\end{equation}
Thus, from Eq.~\eqref{eq:diffusion_constant}, we obtain
\begin{subequations}
	\begin{equation}
		\mathcal{D}_{xx} = v_x^2 \tau^\textrm{qp}_\textrm{F},
	\end{equation}
	\begin{equation}
		\mathcal{D}_{yy} = v_y^2 \tau^\textrm{qp}_\textrm{F}.
	\end{equation}
\end{subequations}

On the other hand, the long-range Coulomb impurity potential within the Thomas-Fermi approximation is given by
\begin{equation}
	V(\bm{q}) = \frac{2\pi e^2}{\epsilon_0 (q + q_\textrm{TF})},
\end{equation}
where $q \equiv |\bm{q}|$, $\epsilon_0$ is the background dielectric constant, and $q_\textrm{TF}$ is the Thomas-Fermi wavevector given by
\begin{equation}
	q_\textrm{TF} = \frac{2 \pi e^2}{\epsilon_0} N(\varepsilon_\textrm{F}) = \frac{4 e^2 \varepsilon_\textrm{F}}{\epsilon_0 v_x v_y}.
\end{equation}
Inserting $W_{\bm{k};\bm{k}^\prime} = 2\pi n_\textrm{imp} \left| V(\bm{q}) \right|^2 F_{\bm{k}\bm{k}^\prime} \delta(\varepsilon-\varepsilon^\prime)$ into Eq.~\eqref{eq:quasiparticle_lifetime}, the quasiparticle lifetime at the Fermi energy is given by
\begin{equation}
	\label{eq:graphene_tau_qp_long}
	\frac{\tau_0}{\tau^\textrm{qp} (\varepsilon_\textrm{F},\theta)} = \int_0^{2\pi} d\theta^\prime \frac{1 + \cos(\theta - \theta^\prime)}{\left[ \tilde{q} (\theta, \theta^\prime) + Q \right]^2}.
\end{equation}
Here, $\tau_0 \equiv \varepsilon_\textrm{F}/\varepsilon_0^2$, $Q \equiv q_\textrm{TF}/k_\textrm{F}$, and $\tilde{q}(\theta,\theta^\prime)$ is defined by
\begin{equation}
	\tilde{q}(\theta,\theta^\prime) \equiv \left[ A (\cos\theta - \cos\theta^\prime)^2 + \frac{1}{A} (\sin\theta - \sin\theta^\prime)^2  \right]^{\frac{1}{2}},
\end{equation}
where $\varepsilon_0 \equiv \frac{e^2}{\epsilon_0 d_0}$, $d_0 \equiv (\pi n_\textrm{imp})^{-\frac{1}{2}}$ is the average distance between impurities, $A \equiv k_{\textrm{F}}^{(x)} / k_{\textrm{F}}^{(y)} = v_y / v_x$ is the anisotropy factor, $k_{\textrm{F}}^{(i)}$ is the Fermi wavevector along the $i$-th direction, and $k_\textrm{F}$ is the effective Fermi wavevector defined by mapping $\pi k_\textrm{F}^2$ to the area inside the Fermi surface so that
\begin{equation}
	\pi k_\textrm{F}^2 = \frac{\pi \varepsilon_\textrm{F}^2}{v_x v_y}.
\end{equation}
On the other hand, inserting $W_{\bm{k};\bm{k}^\prime}$ into Eq.~\eqref{eq:transport_relaxation_time}, we obtain
\begin{subequations}
	\begin{equation}
		\label{eq:graphene_tau_x_long}
		\left[ \frac{\tau^{(x)} (\varepsilon_\textrm{F},\theta)}{\tau^\textrm{qp} (\varepsilon_\textrm{F},\theta)} - 1 \right] \cos\theta = \frac{1}{\tau_0} \int_0^{2\pi} d\theta^\prime \frac{1 + \cos(\theta - \theta^\prime)}{\left[ \tilde{q} (\theta, \theta^\prime) + Q \right]^2} \tau^{(x)} (\varepsilon_\textrm{F},\theta^\prime) \cos\theta^\prime,
	\end{equation}
	\begin{equation}
		\label{eq:graphene_tau_y_long}
		\left[ \frac{\tau^{(y)} (\varepsilon_\textrm{F},\theta)}{\tau^\textrm{qp} (\varepsilon_\textrm{F},\theta)} - 1 \right] \sin\theta = \frac{1}{\tau_0} \int_0^{2\pi} d\theta^\prime \frac{1 + \cos(\theta - \theta^\prime)}{\left[ \tilde{q} (\theta, \theta^\prime) + Q \right]^2} \tau^{(y)} (\varepsilon_\textrm{F},\theta^\prime) \sin\theta^\prime.
	\end{equation}
\end{subequations}
Now, we write the transport relaxation time as
\begin{equation}
	\frac{\tau^{(i)} (\varepsilon_\textrm{F}, \theta)}{\tau_0} = \sum_n a_n^{(i)} \cos(2n\theta),
\end{equation}
where $i = x,y$ and $n=0,1,2,\cdots$. Note that only $2n$ factors are possible by the assumption $\varepsilon(\bm{k}) = \varepsilon(-\bm{k})$. Inserting a distinct set of angles $\{ \theta_m \}$ into Eqs.~\eqref{eq:graphene_tau_x_long} and \eqref{eq:graphene_tau_y_long}, we obtain the following linear equations:
\begin{subequations}
	\label{eq:graphene_relaxation_time_matrix_equation}
	\begin{equation}
		\sum_n M_{mn}^{(x)} a_n^{(x)} = \cos\theta_m,
	\end{equation}
	\begin{equation}
		\sum_n M_{mn}^{(y)} a_n^{(y)} = \sin\theta_m,
	\end{equation}
\end{subequations}
where the matrix elements $M_{mn}^{(x,y)}$ are given by
\begin{subequations}
	\begin{equation}
		M_{mn}^{(x)} = \frac{\cos(2n\theta_m)\cos\theta_m}{\tau^\textrm{qp} (\varepsilon_\textrm{F}, \theta_m) / \tau_0} - \int_0^{2\pi} d\theta^\prime \frac{\cos\theta^\prime \cos(2n\theta^\prime) \left[ 1 + \cos(\theta_m - \theta^\prime) \right]}{\left[ \tilde{q} (\theta_m , \theta^\prime) + Q \right]^2},
	\end{equation}
	\begin{equation}
		M_{mn}^{(y)} = \frac{\cos(2n\theta_m)\sin\theta_m}{\tau^\textrm{qp} (\varepsilon_\textrm{F}, \theta_m) / \tau_0} - \int_0^{2\pi} d\theta^\prime \frac{\sin\theta^\prime \cos(2n\theta^\prime) \left[ 1 + \cos(\theta_m - \theta^\prime) \right]}{\left[ \tilde{q} (\theta_m , \theta^\prime) + Q \right]^2}.
	\end{equation}
\end{subequations}
Hence, we can obtain $a_n^{(i)}$ by solving Eq.~\eqref{eq:graphene_relaxation_time_matrix_equation} with a large enough cutoff for $m$ and $n$. Finally, from Eq.~\eqref{eq:diffusion_constant}, we have
\begin{subequations}
	\label{eq:diffusion_constant_graphene_numerical}
	\begin{equation}
		\mathcal{D}_{xx} = \frac{\mathcal{D}_0}{2\pi A} \int_0^{2\pi} d\theta \cos^2 \theta \left[ \sum_n a_n^{(x)} \cos(2n\theta) \right],
	\end{equation}
	\begin{equation}
		\mathcal{D}_{yy} = \frac{\mathcal{D}_0 A}{2\pi} \int_0^{2\pi} d\theta \sin^2 \theta \left[ \sum_n a_n^{(y)} \cos(2n\theta) \right],
	\end{equation}
\end{subequations}
where $\mathcal{D}_0 \equiv v_x v_y \tau_0$.

\subsection{Few-layer black phosphorous}
\label{sec:diffusion_constants_calculation_BP}
The low-energy effective Hamiltonian for few-layer black phosphorus (fBP) at the semi-Dirac transition point is given by \cite{S_Park2019}
\begin{equation}
	H = \frac{k_x^2}{2m^*} \sigma_x + v_0 k_y \sigma_y,
\end{equation}
where $m^*$ is the effective mass along the zigzag ($x$) direction, $v_0$ is the velocity along the armchair ($y$) direction, and $\bm{\sigma}$ is a vector of Pauli matrices. Let us consider the coordinate transformation $(k_x , k_y) = (\eta \sqrt{2m^* \varepsilon\cos\theta} , \frac{\varepsilon}{v_0} \sin\theta)$, where $\eta = \pm 1$ represents each half of $\bm k$ space and $\theta$ varies from $-\pi/2$ to $\pi/2$, with the Jacobian $\mathcal{J} (\eta,\varepsilon,\theta) = \frac{1}{v_0} \sqrt{\frac{m^* \varepsilon}{2\cos\theta}}$. Then, the Hamiltonian transforms into $H = \varepsilon (\cos\theta\sigma_x + \sin\theta\sigma_y)$ and its eigenvalues and eigenvectors are given by $E (\eta,\varepsilon,\theta) = \varepsilon$ and $\big| \eta, \varepsilon, \theta \big> = \frac{1}{\sqrt{2}} (1, e^{i\theta})^{\rm{t}}$ ensuring that the overlap factor is given by $F_{\bm{k}\bm{k}^\prime} = \frac{1}{2} \left[ 1 + \cos(\theta - \theta^\prime) \right]$. Here, we assume $\varepsilon > 0$. Also, note that the $x$ and $y$ components of the velocity are given by $v^{(x)} (\eta, \varepsilon,\theta) = \eta \sqrt{\frac{2\varepsilon}{m^*}} \cos^{3/2} \theta$ and $v^{(y)} (\eta, \varepsilon,\theta) = v_0 \sin\theta$, respectively, and the density of states is given by
\begin{equation}
	N(\varepsilon) = \frac{\mathrm{g} F(\pi/4, 2) \sqrt{2 m^* \varepsilon}}{\pi^2 v_0}.
\end{equation}
where $\mathrm{g} = 2$ is the spin degeneracy factor and $F(\phi, k) = \int_0^\phi d\theta [1 - k\sin^2 \theta]^{-1/2}$ is the elliptic integral of the first kind \cite{S_Boas2006} with $F(\pi/4, 2) \approx 1.311$.

The short-range impurity potential is given by $V(\bm{q}) = V_0$ in the momentum space, where $\bm{q} \equiv \bm{k} - \bm{k}^\prime$. Inserting the transition rate $W_{\bm{k};\bm{k}^\prime} = 2\pi n_\textrm{imp} \left| V(\bm{q}) \right|^2 F_{\bm{k}\bm{k}^\prime} \delta(\varepsilon-\varepsilon^\prime)$ into Eq.~\eqref{eq:quasiparticle_lifetime}, the quasiparticle lifetime at the Fermi energy $\varepsilon_\textrm{F}$ is given by
\begin{align}
	\frac{1}{\tau^\textrm{qp} (\varepsilon_\textrm{F}, \theta)} = \frac{1}{\tau_\textrm{imp}} \int_{-\pi/2}^{\pi/2} d\theta^\prime \frac{1 + \cos(\theta - \theta^\prime)}{\sqrt{\cos\theta^\prime}} = \frac{4}{\tau_\textrm{imp}} \left[ F \left( \frac{\pi}{4}, 2 \right) + E \left( \frac{\pi}{4}, 2 \right) \cos\theta \right],
\end{align}
where $E(\phi, k) = \int_0^\phi d\theta [1 - k\sin^2 \theta]^{1/2}$ is the elliptic integral of the second kind \cite{S_Boas2006} with $E(\pi/4, 2) \approx 0.5991$, and $\tau_\textrm{imp}$ is defined by
\begin{equation}
	\frac{1}{\tau_\textrm{imp}} \equiv \frac{n_\textrm{imp} V_0^2}{2\pi v_0} \sqrt{\frac{m^* \varepsilon_\textrm{F}}{2}}.
\end{equation}
On the other hand, inserting $W_{\bm{k};\bm{k}^\prime}$ into Eq.~\eqref{eq:transport_relaxation_time}, we obtain the following integral equations for the transport relaxation time:
\begin{subequations}
	\begin{equation}
		\label{eq:BP_tau_x}
		\left[ \frac{\tau^{(x)} (\eta, \varepsilon_\textrm{F}, \theta)}{\tau^\textrm{qp} (\varepsilon_\textrm{F}, \theta)} - 1 \right] \eta \cos^{3/2} \theta = \sum_{\eta^\prime = \pm 1} \eta^\prime \int_{-\pi/2}^{\pi/2} d\theta^\prime \frac{\tau^{(x)}(\eta^\prime, \varepsilon_\textrm{F}, \theta^\prime)}{\tau_\textrm{imp}} \frac{\cos\theta^\prime \left[ 1 + \cos(\theta-\theta^\prime)\right]}{2},
	\end{equation}
	\begin{equation}
		\label{eq:BP_tau_y}
		\left[ \frac{\tau^{(y)} (\eta, \varepsilon_\textrm{F}, \theta)}{\tau^\textrm{qp} (\varepsilon_\textrm{F}, \theta)} - 1 \right] \sin \theta = \sum_{\eta^\prime = \pm 1} \int_{-\pi/2}^{\pi/2} d\theta^\prime \frac{\tau^{(y)}(\eta^\prime, \varepsilon_\textrm{F}, \theta^\prime)}{\tau_\textrm{imp}} \frac{\sin\theta^\prime \left[ 1 + \cos(\theta - \theta^\prime)\right]}{2\sqrt{\cos\theta^\prime}}.
	\end{equation}
\end{subequations}
Since the transport relaxation times are independent of $\eta$ by the reflection symmetric dispersion about the $k_y$ direction, the right-hand side of Eq.~\eqref{eq:BP_tau_x} vanishes so that
\begin{equation}
	\tau^{(x)} (\varepsilon_\textrm{F},\theta) = \tau^\textrm{qp} (\varepsilon_\textrm{F},\theta).
\end{equation}
On the other hand, expanding $\cos(\theta - \theta^\prime) = \cos\theta \cos\theta^\prime + \sin\theta \sin\theta^\prime$, Eq.~\eqref{eq:BP_tau_y} transforms into
\begin{equation}
	\label{eq:BP_tau_ratio}
	\frac{\tau^{(y)} (\eta, \varepsilon_\textrm{F}, \theta)}{\tau^\textrm{qp} (\varepsilon_\textrm{F}, \theta)} - 1 = \int_{-\pi/2}^{\pi/2} d\theta^\prime \frac{\tau^{(y)}(\eta^\prime, \varepsilon_\textrm{F}, \theta^\prime)}{\tau_\textrm{imp}} \frac{\sin^2 \theta^\prime}{\sqrt{\cos\theta^\prime}}.
\end{equation}
Here, we use that $\tau^{(y)} (\varepsilon_\textrm{F}, \theta)$ is an even function of $\theta$ by the reflection symmetric dispersion about the $k_x$ direction. Considering that the right-hand side of Eq.~\eqref{eq:BP_tau_ratio} is independent of $\theta$, we can numerically obtain $S \equiv \tau^{(y)} (\varepsilon_\textrm{F}, \theta) / \tau^\textrm{qp} (\varepsilon_\textrm{F}, \theta)$ given by
\begin{align}
	S & = 1 + \frac{S}{4} \int_{-\pi/2}^{\pi/2} d\theta^\prime \frac{1}{F(\pi/4, 2) + E(\pi/4, 2) \cos\theta^\prime} \frac{\sin^2 \theta^\prime}{\sqrt{\cos\theta^\prime}}  = 2.491.
\end{align}
Thus, from Eq.~\eqref{eq:diffusion_constant}, we obtain
\begin{subequations}
	\begin{equation}
		\mathcal{D}_{xx} = \frac{\varepsilon_\textrm{F} \tau_\textrm{imp}}{8F(\pi/4,2) m^*} \int_{-\pi/2}^{\pi/2} d\theta \frac{\cos^{5/2} \theta}{F(\pi/4,2) + E(\pi/4,2) \cos\theta} = 7.507 \times 10^{-2} \frac{\varepsilon_\textrm{F}}{m^*} \tau_\textrm{imp},
	\end{equation}
	\begin{equation}
		\mathcal{D}_{yy} = \frac{Sv_0^2 \tau_\textrm{imp}}{16F(\pi/4,2)} \int_{-\pi/2}^{\pi/2} d\theta \frac{\sin^{2} \theta}{\left[ F(\pi/4,2) + E(\pi/4,2) \cos\theta \right] \sqrt{\cos\theta}} = 0.2844 v_0^2 \tau_\textrm{imp}.
	\end{equation}
\end{subequations}

On the other hand, the long-range Coulomb impurity potential within the Thomas-Fermi approximation is given by 
\begin{equation}
	V(\bm{q}) = \frac{2\pi e^2}{\epsilon_0 (q + q_\textrm{TF})},
\end{equation}
where $q \equiv |\bm{q}|$, $\epsilon_0$ is the background dielectric constant constant, and $q_\textrm{TF}$ is the Thomas-Fermi wavevector given by
\begin{equation}
	q_\textrm{TF} = \frac{2\pi e^2}{\epsilon_0} N(\varepsilon_\textrm{F}) = \frac{2F(\pi/4, 2) \mathrm{g}e^2}{\pi \epsilon_0 v_0} \sqrt{2m^* \varepsilon_\textrm{F}}.
\end{equation}
Inserting $W_{\bm{k};\bm{k}^\prime} = 2\pi n_\textrm{imp} \left| V(\bm{q}) \right|^2 F_{\bm{k}\bm{k}^\prime} \delta(\varepsilon-\varepsilon^\prime)$ into Eq.~\eqref{eq:quasiparticle_lifetime}, the quasiparticle lifetime at the Fermi energy is given by
\begin{equation}
	\frac{\tau_0}{\tau^\textrm{qp} (\varepsilon_\textrm{F}, \theta)} = \sum_{\eta^\prime = \pm 1} \int_{-\pi/2}^{\pi/2} d\theta^\prime \frac{1 + \cos(\theta - \theta^\prime)}{\left[ \tilde{q} (\theta, \eta^\prime, \theta^\prime) + \mathfrak{C} Q \right]^2 \sqrt{\cos\theta^\prime}}.
\end{equation}
Here, $\tau_0 \equiv 2\varepsilon_\textrm{F}/\varepsilon_0^2$, $\mathfrak{C} = \left[ \frac{8F(\pi/4, 2)}{3\pi} \right]^{1/2} \approx 1.055$, $Q \equiv q_\textrm{TF}/k_\textrm{F}$, and $\tilde{q} (\theta, \eta^\prime, \theta^\prime)$ is defined by
\begin{equation}
	\tilde{q} (\theta, \eta^\prime, \theta^\prime) \equiv \bigg[ A \left( \sqrt{\cos\theta} - \eta^\prime \sqrt{\cos\theta^\prime} \right)^2 + \frac{1}{A} \left( \sin\theta - \sin\theta^\prime \right)^2 \bigg]^{1/2},
\end{equation}
where $\varepsilon_0 \equiv \frac{e^2}{\epsilon_0 d_0}$, $d_0 \equiv (\pi n_\textrm{imp})^{-\frac{1}{2}}$ is the average distance between impurities, $A \equiv k_{\textrm{F}}^{(x)} / k_{\textrm{F}}^{(y)} = \sqrt{2m^* v_0^2 / \varepsilon_\textrm{F}}$ is the anisotropy factor, $k_{\textrm{F}}^{(i)}$ is the Fermi wavevector along the $i$-th direction, and $k_\textrm{F}$ is the effective Fermi wavevector defined by mapping $\pi k_\textrm{F}^2$ to the area inside the Fermi surface so that
\begin{equation}
	\pi k_\textrm{F}^2 = \frac{8F(\pi/4, 2)}{3v_0} \sqrt{2m^* \varepsilon_\textrm{F}^3}.
\end{equation}
Note that we choose $\eta = 1$ since the quasiparticle lifetime is independent of $\eta$ by the reflection symmetric dispersion about the $k_y$ axis. On the other hand, inserting $W_{\bm{k};\bm{k}^\prime}$ into Eq.~\eqref{eq:transport_relaxation_time}, we obtain
\begin{subequations}
	\begin{equation}
		\label{eq:BP_tau_x_long}
		\left[ \frac{\tau^{(x)} (\varepsilon_\textrm{F}, \theta)}{\tau^\textrm{qp} (\varepsilon_\textrm{F}, \theta)} - 1 \right] \cos^{3/2} \theta = \sum_{\eta^\prime = \pm 1} \eta^\prime \int_{-\pi/2}^{\pi/2} d\theta^\prime \frac{ \left[ 1 + \cos(\theta - \theta^\prime) \right] \cos\theta^\prime}{\left[ \tilde{q} (\theta, \eta^\prime, \theta^\prime) + \mathfrak{C} Q \right]^2} \frac{\tau^{(x)} (\varepsilon_\textrm{F}, \theta^\prime)}{\tau_0}, \nonumber
	\end{equation}
	\begin{equation}
		\label{eq:BP_tau_y_long}
		\left[ \frac{\tau^{(y)} (\varepsilon_\textrm{F}, \theta)}{\tau^\textrm{qp} (\varepsilon_\textrm{F}, \theta)} - 1 \right] \sin \theta = \sum_{\eta^\prime = \pm 1} \int_{-\pi/2}^{\pi/2} d\theta^\prime \frac{ \left[ 1 + \cos(\theta - \theta^\prime) \right] \sin\theta^\prime}{\left[ \tilde{q} (\theta, \eta^\prime, \theta^\prime) + \mathfrak{C} Q \right]^2 \sqrt{\cos\theta^\prime}} \frac{\tau^{(y)} (\varepsilon_\textrm{F}, \theta^\prime)}{\tau_0}. \nonumber
	\end{equation}
\end{subequations}
Now, we write the transport relaxation time as
\begin{equation}
	\frac{\tau^{(i)} (\varepsilon_\textrm{F}, \theta)}{\tau_0} = \sum_n a_n^{(i)} \cos(n\theta),
\end{equation}
where $i=x,y$ and $n=0,1,2,\cdots$. Here, only cosine series is possible by the reflection symmetric dispersion about the $k_y$ axis. Inserting a distinct set of angles $\{ \theta_m \}$ into Eqs.~\eqref{eq:BP_tau_x_long} and \eqref{eq:BP_tau_y_long}, we obtain the following linear equations:
\begin{subequations}
	\label{eq:BP_relaxation_time_matrix_equation}
	\begin{equation}
		\sum_n M_{mn}^{(x)} a_n^{(x)} = \cos^{\frac{3}{2}} \theta_m,
	\end{equation}
	\begin{equation}
		\sum_n M_{mn}^{(y)} a_n^{(y)} = \sin\theta_m,
	\end{equation}
\end{subequations}
where the matrix elements $M_{mn}^{(x,y)}$ are given by
\begin{subequations}
	\begin{equation}
		M_{mn}^{(x)} = \frac{\cos(n\theta_m)\cos^{3/2}\theta_m}{\tau^\textrm{qp} (\varepsilon_\textrm{F}, \theta_m) / \tau_0} - \sum_{\eta^\prime = \pm 1} \eta^\prime \int_{-\pi/2}^{\pi/2} d\theta^\prime \frac{\cos\theta^\prime \cos(n\theta^\prime) \left[ 1 + \cos(\theta_m - \theta^\prime) \right]}{\left[ \tilde{q} (\theta_m , \eta^\prime, \theta^\prime) + \mathfrak{C} Q \right]^2},
	\end{equation}
	\begin{equation}
		M_{mn}^{(y)} = \frac{\cos(n\theta_m)\sin\theta_m}{\tau^\textrm{qp} (\varepsilon_\textrm{F}, \theta_m) / \tau_0} - \sum_{\eta^\prime = \pm 1} \int_{-\pi/2}^{\pi/2} d\theta^\prime \frac{\sin\theta^\prime \cos(n\theta^\prime) \left[ 1 + \cos(\theta_m - \theta^\prime) \right]}{\left[ \tilde{q} (\theta_m , \eta^\prime, \theta^\prime) + \mathfrak{C} Q \right]^2 \sqrt{\cos\theta^\prime}}.
	\end{equation}
\end{subequations}
Hence, we can obtain $a_n^{(i)}$ by solving Eq.~\eqref{eq:BP_relaxation_time_matrix_equation} with a large enough cutoff for $m$ and $n$. Finally, from Eqs.~\eqref{eq:diffusion_constant}, we have
\begin{subequations}
	\label{eq:diffusion_constant_BP_numerical}
	\begin{equation}
		\mathcal{D}_{xx} = \frac{2\mathcal{D}_0}{A}  \int_{-\pi/2}^{\pi/2} d\theta \cos^{5/2} \theta \left[ \sum_n a_n^{(x)} \cos(n\theta) \right],
	\end{equation}
	\begin{equation}
		\mathcal{D}_{yy} = \frac{\mathcal{D}_0 A}{2} \int_{-\pi/2}^{\pi/2} d\theta \frac{\sin^2 \theta}{\sqrt{\cos\theta}} \left[ \sum_n a_n^{(y)} \cos(n\theta) \right],
	\end{equation}
\end{subequations}
where $\mathcal{D}_0$ is defined by
\begin{equation}
	\mathcal{D}_0 \equiv \frac{v_0 \tau_0}{4F(\pi/4, 2)} \sqrt{\frac{2\varepsilon_\textrm{F}}{m^*}}.
\end{equation}

\subsection{Anisotropy of the diffusion constant and conductivity}
For anisotropic 2DEG, anisotropic graphene, and fBP at the semi-Dirac transition point, we calculate the ratio between $\mathcal{D}_{yy} / \mathcal{D}_{xx}$ and its commonly expected value $(v_{\textrm{F}}^{(y)} / v_{\textrm{F}}^{(x)})^2$,
assuming the short-range and long-range disorders, respectively. For short-range disorder, we check whether the ratio $(\mathcal{D}_{yy} / \mathcal{D}_{xx}) / (v_{\textrm{F}}^{(y)} / v_{\textrm{F}}^{(x)})^2$ coincides with $1$. For long-range disorder, we numerically calculate and plot $(\mathcal{D}_{yy} / \mathcal{D}_{xx}) / (v_{\textrm{F}}^{(y)} / v_{\textrm{F}}^{(x)})^2$ as a function of the screening factor $Q \equiv q_\textrm{TF} / k_\textrm{F}$ assuming $A = 2,5$, and also as a function of the anisotropy factor $A$ assuming $Q = 0.1, 5$, and analyze its deviation from $1$. 
%Note that values of $A$ and $Q$ used in the calculation for fBP at the semi-Dirac transition point are estimated from the realistic values in tri-layer and tetra-layer black phosphorus (3BP and 4BP, respectively), whereas for anisotropic 2DEG and anisotropic graphene $A$ and $Q$ are set equal to that in fBP for comparison. 
Note that values of $A$ and $Q$ used in the calculation 
are estimated from fBP at the semi-Dirac transition point with realistic parameters for trilayer and tetralayer black phosphorus (3BP and 4BP, respectively). For anisotropic 2DEG and anisotropic graphene, we use the same $A$ and $Q$ for comparison. 

%In detail, the realistic parameters in fBP are known as $m^* = 1.061 m_e$ and $v_0 = 1.486 \times 10^7$ $\mathrm{cm/s}$ for 3BP and $m^* = 0.930 m_e$ and $v_0 = 1.246 \times 10^7$ $\mathrm{cm/s}$ for 4BP, where $m_e$ is the electron mass \cite{S_Jang2019}. Since the doping concentration in two-dimensional systems typically ranges from $10^{12}$ to $10^{13}$ $\mathrm{cm}^{-2}$, $A$ ranges from 2.37 to 5.10 for 3BP and from 1.93 to 4.15 for 4BP. Thus, we choose $A=2$ and $A=5$ which are approximately the smallest and largest possible values of $A$, respectively. On the other hand, assuming $\epsilon_0 \approx 10$ (dielectric constant of SiC substrate), we choose $Q=5$ for the calculation in the range from 3.87 to 5.67 for 3BP and 4.16 to 6.11 for 4BP. Note that $Q=0.1$ which would be unrealistic is additionally chosen for convenience in comparison.
In detail, the realistic parameters in fBP are known as $m^* = 1.061 m_e$ and $v_0 = 1.486 \times 10^7$ $\mathrm{cm/s}$ for 3BP and $m^* = 0.930 m_e$ and $v_0 = 1.246 \times 10^7$ $\mathrm{cm/s}$ for 4BP, where $m_e$ is the electron mass \cite{S_Jang2019}. For a typical doping concentration $n = 10^{12}-10^{13}$ $\mathrm{cm}^{-2}$ in 2D systems including fBP-based devices \cite{S_Xia2014, S_Xiang2015}, $A$ is given by 5.10 (4.15) for 3BP (4BP) at $n = 10^{12}$ $\mathrm{cm}^{-2}$ and 2.37 (1.93) for 3BP (4BP) at $n = 10^{13}$ $\mathrm{cm}^{-2}$. 
%Thus, we choose $A=2$ and $A=5$ which are approximately the smallest and largest possible values of $A$, respectively. 

For short-range disorder, we obtain $(\mathcal{D}_{yy} / \mathcal{D}_{xx}) / (v_{\textrm{F}}^{(y)} / v_{\textrm{F}}^{(x)})^2 = 1$ for anisotropic 2DEG and anisotropic graphene, and $(\mathcal{D}_{yy} / \mathcal{D}_{xx}) / (v_{\textrm{F}}^{(y)} / v_{\textrm{F}}^{(x)})^2 \approx 7.58$ for fBP. From the result of anisotropic graphene, we infer that the chiral wave function of the system does not generate the deviation of $(\mathcal{D}_{yy} / \mathcal{D}_{xx}) / (v_{\textrm{F}}^{(y)} / v_{\textrm{F}}^{(x)})^2$ from 1. 
However, when the system has a different power-law dispersion in each direction as in fBP at the semi-Dirac transition point, $(\mathcal{D}_{yy} / \mathcal{D}_{xx}) / (v_{\textrm{F}}^{(y)} / v_{\textrm{F}}^{(x)})^2$ shows a considerable derivation from 1 even for short-range disorder.

\begin{figure*}[htb]
	\includegraphics[width=1.0\linewidth]{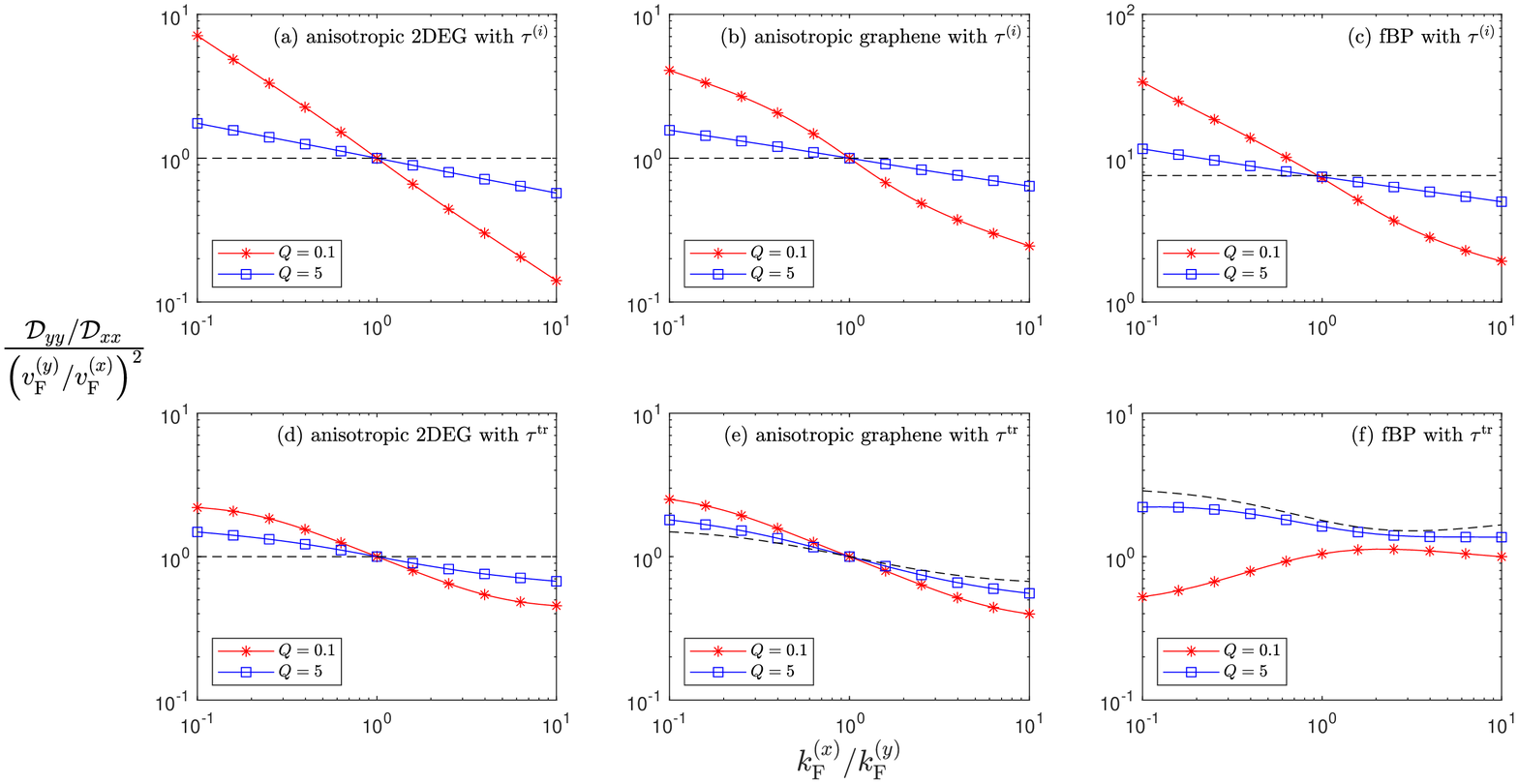}
	\caption{Ratio $(\mathcal{D}_{yy} / \mathcal{D}_{xx}) / (v_{\textrm{F}}^{(y)} / v_{\textrm{F}}^{(x)})^2$ as a function of the anisotropy factor $A = k_{\textrm{F}}^{(x)} / k_{\textrm{F}}^{(y)}$ assuming $Q=0.1, 5$ for (a), (c) anisotropic 2DEG, (b), (e) anisotropic graphene, and (c), (f) fBP at the semi-Dirac transition point, obtained from (a)-(c) Eq.~\eqref{eq:diffusion_constant} and from (d)-(e) Eq.~\eqref{eq:relaxation_time_approximation} as in isotropic systems. The values for the short-range disorder are represented by the black dashed lines.} 
	\label{fig:charged_impurity_A_plot}
\end{figure*}

For long-range disorder as illustrated in Figs.~\ref{fig:diffusion_anisotropy}(a)-\ref{fig:diffusion_anisotropy}(c) we observe that $(\mathcal{D}_{yy} / \mathcal{D}_{xx}) / (v_{\textrm{F}}^{(y)} / v_{\textrm{F}}^{(x)})^2$ decreases from the short-range disorder result as the screening becomes weaker, approaching the short-range disorder result in the strong screening result. The deviation from the short-range disorder result also increases as the anisotropy of the system increases, as shown in Figs.~\ref{fig:charged_impurity_A_plot}(a)-\ref{fig:charged_impurity_A_plot}(c) which presents the dependence on the anisotropy for given screening strength $Q$. (Here, assuming $\epsilon_0 \approx 10$ for the dielectric constant of SiC substrate, we choose $Q=5$ for the calculation as well as $Q=0.1$ for comparison.)
This indicates that the difference between the transport relaxation times $\tau^{(x)}$ and $\tau^{(y)}$ becomes significant as the anisotropy of the system increases or the screening becomes weaker. Also note that the deviation of $(\mathcal{D}_{yy} / \mathcal{D}_{xx}) / (v_{\textrm{F}}^{(y)} / v_{\textrm{F}}^{(x)})^2$ from the short-range disorder result shows a stronger dependence on both the screening strength and anisotropy of the system compared to that obtained from the relaxation time neglecting the component-dependence given by Eq.~\eqref{eq:relaxation_time_approximation} as in isotropic systems [Figs.~\ref{fig:diffusion_anisotropy}(d)-\ref{fig:diffusion_anisotropy}(f) and Figs.~\ref{fig:charged_impurity_A_plot}(d)-\ref{fig:charged_impurity_A_plot}(f)].

\end{document}